\documentclass[twocolumn]{aastex63}

\usepackage{graphicx}
\usepackage{graphics}
\usepackage{amssymb}
\usepackage{bm}
\usepackage{times}
\usepackage[flushleft]{threeparttable}
\usepackage{booktabs}
\usepackage{footnote}
\usepackage{multirow}
\usepackage{color}

\newcommand{\kms}{km~s$^{-1}$}

\newcommand{\ha}{\ensuremath{{\rm H}\alpha}}

\newcommand{\arcs}{\ensuremath{^{\prime\prime}}}
\definecolor{cerulean}{rgb}{0.0, 0.48, 0.65}
\definecolor{blue}{rgb}{0.0, 0.0, 1.0}
\definecolor{red}{rgb}{1.0, 0.0, 0.0}

\newcommand{\hicm}{H{\sc i} 21 cm }
\newcommand{\hi}{H{\sc i} }

\shorttitle{Local starburst conditions in  GRB 980425 host}
\shortauthors{Arabsalmani et al.}
\begin{document}
\title%{Molecular gas clumps   in the metal-poor environment of a massive star explosion}
%{Molecular gas clumps and starburst conditions in the host galaxy of GRB 980425 / SN 1998bw} 
{Local starburst conditions  and formation of  GRB 980425 / SN 1998bw within a collisional ring}
%{Local starburst conditions  in the host galaxy of   GRB 980425 / SN 1998bw}
%{Formation of a GRB progenitor triggered by starburst conditions in a collisional ring}

\correspondingauthor{Maryam Arabsalmani}
\email{maryam.arabsalmani@icrar.org}

\author{M. Arabsalmani}, 
\altaffiliation{}
\affiliation{International Centre for Radio Astronomy Research (ICRAR), M468, University of Western Australia, 35 Stirling Hwy, Crawley, WA 6009, Australia}
\affiliation{ARC Centre of Excellence for All Sky Astrophysics in 3 Dimensions (ASTRO 3D), Australia}

\author{F. Renaud}
\affiliation{Department of Astronomy and Theoretical Physics, Lund Observatory, Box 43, 221 00 Lund, Sweden}

\author{S. Roychowdhury}
\affiliation{International Centre for Radio Astronomy Research (ICRAR), M468, University of Western Australia, 35 Stirling Hwy, Crawley, WA 6009, Australia}
\affiliation{ARC Centre of Excellence for All Sky Astrophysics in 3 Dimensions (ASTRO 3D), Australia}

\author{V. Arumugam}
\affiliation{Institut de Radioastronomie Millim\'etrique, 300 rue de la Piscine, Domaine Universitaire, 38406 Saint Martin d'H\'eres, France}

\author{E. Le Floc'h}
\affiliation{IRFU, CEA, Universit\'e Paris-Saclay, F-91191 Gif-sur-Yvette, France}
\affiliation{Universit\'e Paris Diderot, AIM, Sorbonne Paris Cit\'e, CEA, CNRS, F-91191 Gif-sur-Yvette, France}

\author{F. Bournaud}
\affiliation{IRFU, CEA, Universit\'e Paris-Saclay, F-91191 Gif-sur-Yvette, France}
\affiliation{Universit\'e Paris Diderot, AIM, Sorbonne Paris Cit\'e, CEA, CNRS, F-91191 Gif-sur-Yvette, France}

\author{D. Cormier}
\affiliation{IRFU, CEA, Universit\'e Paris-Saclay, F-91191 Gif-sur-Yvette, France}
\affiliation{Universit\'e Paris Diderot, AIM, Sorbonne Paris Cit\'e, CEA, CNRS, F-91191 Gif-sur-Yvette, France}

\author{M. A. Zwaan}
\affiliation{European Southern Observatory, Karl-Schwarzschildstrasse 2, 85748 Garching Bei Muenchen, Germany}

\author{L. Christensen}
\affiliation{DARK, Niels Bohr Institute, University of Copenhagen, Lyngbyvej 2, DK-2100 Copenhagen {\O}, Denmark}

\author{E. Pian}
\affiliation{INAF, Astrophysics and Space Science Observatory, via P. Gobetti 101, 40129 Bologna, Italy}

\author{S. Madden}
\affiliation{IRFU, CEA, Universit\'e Paris-Saclay, F-91191 Gif-sur-Yvette, France}
\affiliation{Universit\'e Paris Diderot, AIM, Sorbonne Paris Cit\'e, CEA, CNRS, F-91191 Gif-sur-Yvette, France}

\author{A. Levan}
\affiliation{Department of Physics, University of Warwick, Coventry, CV4 7AL, UK}

%---------------------------------------------------------------------------------------------------
%---------------------------------------------------------------------------------------------------

\begin{abstract}
We present the first spatially resolved  study  of molecular gas    in the vicinity of  a Gamma Ray Burst, 
using CO(2-1) emission line observations with  the Atacama Large Millimetre Array (ALMA)  at $\sim$50 pc scales. 
The host galaxy of GRB 980425 contains a ring of high column density  \hi 
gas which is likely to have formed due to a collision between the GRB host and its companion galaxy, within which 
the GRB is located.  
We detect eleven  molecular gas   clumps in the galaxy, seven of which   are within the gas ring.  
The clump closest to the GRB position  is at a projected separation of $\sim$280 pc. 
Although it is plausible that the GRB progenitor was ejected from clusters formed in this clump, we argue that the 
in situ formation of the GRB progenitor is the most   likely scenario. 
We  measure  the molecular gas masses of the clumps and find them to be sufficient for forming massive star clusters. 
The  molecular gas depletion times of the clumps  show a variation of $\sim$2 dex, comparable with  the large variation 
in depletion times   found in  starburst galaxies in the nearby Universe.     
This  demonstrates  the presence of   starburst modes of star formation on local scales in the galaxy, even while the galaxy 
as a whole cannot be categorised as a starburst based on its global properties. 
% which  strengthens  the idea of a local enhancement of star formation triggered by the passage of a ring-shaped density wave. % due to a  collision between the galaxy and its companion.  
%The  upper limit on the depletion time at the location of the GRB  supports the scenario in which   the progenitor  of GRB 9802425   originated in a  young massive star cluster formed in the starburst mode of star formation. 
Our findings suggest that the progenitor  of GRB 9802425  was originated in  a young massive star cluster formed in the starburst mode of star formation. \\
\end{abstract}

%---------------------------------------------------------------------------------------------------
%---------------------------------------------------------------------------------------------------


%---------------------------------------------------------------------------------------------------
%---------------------------------------------------------------------------------------------------

\section{Introduction} 
\label{sec:int}
%\textbf{Mention atcow study in the line of high gas densities. }

The life-cycle of young, massive stars  ends with  violent  explosions that are  categorised into 
several populations based on the properties  of the resulting emission. Long-duration Gamma 
Ray Bursts (GRBs)  are amongst  the brightest of these explosions, 
with powerful energy releases that make them detectable back to when the first stars and galaxies 
were formed \citep[$z\gtrsim 8$,][]{Tanvir09}. Their extreme emissions, their localisations within active star-forming regions 
\citep[e.g.,][]{Fruchter06-2006Natur.441..463F}, and the discovery of a  number of them in association with Broad-Lined supernovae 
type Ic \citep[SNe Ic-BL, see][and references therein]{Hjorth12-2012grbu.book..169H} strengthens the link between their progenitors 
and massive stars. 
However, the physical properties of the stellar population that gives rise to these explosions and the interstellar medium 
(ISM) conditions  in which they  form remain poorly constrained. 
Detailed studies of the host galaxies of these bright events, and identifying  the factors that distinguish  
the host galaxies from the general star-forming galaxy population can provide valuable insights into addressing these unknowns.

Extensive studies of GRB hosts over the last couple of decades  demonstrate a dichotomy  between their  
hosts and the general star-forming galaxy population, especially at low redshifts   
\citetext{\citealp[e.g.,][]{Savaglio09-2009ApJ...691..182S}, \citealp{Castroceron10-2010ApJ...721.1919C}, 
\citealp{Schulze15-2015ApJ...808...73S}, \citealp{Kruhler15-2015A&A...581A.125K}, \citealp{Perley16-2016ApJ...817....8P};  
\citealp[but see also][]{Greiner15-2015ApJ...809...76G}, \citealp{Arabsalmani18-2018MNRAS.473.3312A}}.  
The typical low mass and metallicities of GRB host galaxies have been  the most noted   distinguishing characteristics.   
These have been  the basis of the paradigm where low  metallicities are required for formation of GRBs and their progenitors 
\citetext{\citealp[e.g.,][]{Wolf07-2007MNRAS.375.1049W}, \citealp{Nuza07-2007MNRAS.375..665N}, \citealp{Perley16-2016ApJ...817....8P}, 
\citealp{Vergani17-2017A&A...599A.120V}, \citealp[but see also][]{Kocevski09-2009ApJ...702..377K} \citealp[and][]{Campisi11-2011MNRAS.417.1013C}}. 
The low metallicities of the host galaxies  have also been long interpreted as supporting evidence for progenitor models which require very metal-poor stars 
\citep[for such a model see e.g.,][]{Yoon06-2006A&A...460..199Y}, failing   to take 
into account the fact that the host metallicities are much higher than the very low metallicties required by those models.  
However, the  detection of a significant fraction of GRBs in metal-rich host 
galaxies  \citep[e.g.,][also in which metal-poor stars cannot form]{Levesque10-2010ApJ...712L..26L, Perley13-2013ApJ...778..128P, Elliott13-2013A&A...556A..23E, Kruhler15-2015A&A...581A.125K, 
Valeev19-2019GCN.25565....1V},      shows that  low metallicity is unlikely to be a requirement  
for their  formation. 
%[see also][for the  detection of  GRB 190829A, one of the closest known GRBs, in a metal-rich spiral galaxy]

The content and structure of molecular gas, the fuel of star formation, and also the efficiency of star formation in the host galaxies, 
are  potentially important factors which have not been discussed extensively in the literature. 
The first few  studies of the molecular gas content  in  GRB host galaxies reported  a puzzling 
deficit  of molecular gas  in the hosts \citep{Hatsukade14-2014Natur.510..247H, Stanway15-2015ApJ...798L...7S, 
Michalowski16-2016A&A...595A..72M}. 
\citet[][]{Arabsalmani18-2018MNRAS.tmp..190A} argued that this  deficit   was the result of  improper comparisons 
or having assumed too  low a  CO-to-molecular-gas conversion factor for the  low metallicity host galaxies. 
In a recent study \citet[][]{Hatsukade20-2020arXiv200209121H} presented the galaxy-scale properties of molecular gas 
for  14 GRB host galaxies  and found them  to follow the same molecular gas scaling relations as the general  star forming 
galaxy population. They also showed that most of the hosts  at $z\lesssim 1$  have somewhat higher molecular gas fractions  
and/or shorter molecular gas depletion times.  
High  gas fractions, typically found in star-forming galaxies at $z\gtrsim 1.5$ \citep[e.g.,][]{Daddi10-2010ApJ...713..686D, 
Santini14-2014A&A...562A..30S},  are thought to allow for  the fast 
collapse of large amounts of gas due to gravitational instability and result in the formation of massive and dense molecular 
clouds in which massive star clusters can form \citep[][]{Dessauges-Zavadsky18-2018MNRAS.479L.118D}. 
%The short molecular gas depletion times suggest  a  starburst mode of star formation, which also favours the formation of massive star clusters \citep[][]{Renaud18-2018NewAR..81....1R}. 

The typical high star formation rate (SFR)  surface densities and high specific star formation rates (sSFRs) of low redshift GRB hosts 
reported by \citet[][]{Kelly14-2014ApJ...789...23K} suggest the presence of  starburst mode of star formation, in which formation of massive 
star clusters are favoured. Formation of GRB progenitors in massive star clusters  is  compatible with the progenitor    
models that involve  interacting  massive stars  in young and dense star clusters \citep[e.g.,][]{Portegies07-2007Natur.450..388P, 
Pan12-2012MNRAS.423.2203P, Heuvel13-2013ApJ...779..114V, Chrimes20-2020MNRAS.491.3479C}.  
%These  indicate that  GRB host galaxies contain dense  regions with   starburst mode of star formation. 
%This is consistent with the progenitor    models involving multiple massive stars interacting in young and dense star clusters \citep[e.g.,][]{Portegies07-2007Natur.450..388P, Pan12-2012MNRAS.423.2203P, Heuvel13-2013ApJ...779..114V, Chrimes20-2020MNRAS.491.3479C}.  
While the  galaxy-scale  studies of GRB hosts indicate that they contain dense  regions with   starburst mode of star formation,  
spatially resolved studies of molecular gas in the host  galaxies are required to 
investigate whether such regions are present in the vicinities of the GRBs. Note that the gas mass in cloud scales (less than a few hundreds pc) 
is  dominated by molecular gas, and hence     measurements of the molecular gas surface densities and 
depletion times  in the explosion sites  can reveal whether the progenitors were formed in the starburst 
mode of star formation. 
Spatially resolved  studies  are  especially important in cases where  internal/external effects cause local starburst conditions, without 
significantly  impacting   the global properties of the host galaxies.

Here we present the first spatially resolved study of molecular gas and its depletion time  in a GRB host galaxy. 
ESO 184--G82 at $z=0.0086$ is the host galaxy of the  closest known GRB, GRB 980425,   one  of the first GRBs discovered in 
association with a  SNe IC-BL \citep[SN1998bw,][]{Galama98-1998Natur.395..670G}. It is consequently one of the 
best-studied host galaxies  \citep[e.g.,][]{Fynbo00, Hammer06-2006A&A...454..103H, Christensen08-2008A&A...490...45C, LeFloch12-2012ApJ...746....7L, Arabsalmani15-2015MNRAS.454L..51A, Kruhler17-2017A&A...602A..85K}. 
%It  follows the Schmit-Kennicutt relation of the normal star-forming galaxies in the nearby Universe.
%The galaxy-scale properties of ESO 184--G82 are similar to those of nearby dwarf galaxies. 
With a  stellar mass of $10^{8.7}\,M_{\odot}$ \citep[][]{Michalowski14-2014A&A...562A..70M} and a SFR of $0.22$ $M_{\odot}\,\rm yr^{-1}$ 
\citep[based on \ha\, emission line and corrected for dust extinction,][]{Kruhler17-2017A&A...602A..85K},  ESO 184--G82  is on the Main Sequence relation in the M$_*$--SFR plane \citep[with  a sSFR of 0.44 Gyr$^{-1}$; see][]{Brinchmann04-2004MNRAS.351.1151B}. 
\citet[][]{Michalowski18-2018A&A...617A.143M} performed  Atacama Pathfinder EXperiment observations of this GRB host and obtained  a brightness temperature luminosity  of $10^{6.67}$ K\,\kms\,pc$^2$ for  its  CO(2-1) emission line. 
From this we calculate   a molecular gas mass  of  $10^{7.8-8.4}\,M_{\odot}$ for ESO 184--G82 (see Section \ref{sec:res} for assumptions used 
when converting the $L^T_{\rm CO(2-1)}$ to molecular gas mass),  similar to those of nearby galaxies with similar stellar masses \citep[see Figure 5 of][]{Grossi16-2016A&A...590A..27G}.

With  an  atomic gas mass of $10^{9.0}\,M_{\odot}$, obtained from \hicm emission line observation \citep[][]{Arabsalmani15-2015MNRAS.454L..51A}, 
ESO 184--G82 has a total gas (atomic+molecular)  fraction of $\sim$2, and a total gas depletion time of $\sim$5 Gyr,  
{
larger than the typical depletion times of starburst 
galaxies in the nearby Universe \citep[see Fig.5 of][]{Kennicutt98-1998ApJ...498..541K}. 
In fact the gas and star formation surface densities of ESO 184--G82, $\Sigma_{\rm gas} \sim 14\,M_{\odot}\rm\,pc^{-2}$    and 
$\Sigma_{\rm SFR} \sim 0.003\,M_{\odot}\,\rm yr^{-1}\,kpc^{-2}$, place it amongst the non-starbursting galaxies  in  the Kennicutt-Schmidt relation 
\citetext{\citealp[see Figure 2 of][]{Kennicutt98-1998ApJ...498..541K}, \citealp[and Figure 6 of][]{delosReyes19-2019ApJ...872...16D}}. 
%, \citealp[see also][]{Roychowdhury11-2011MNRAS.414L..55R}
The surface densities are corrected for the inclination angle of 50$^{\circ}$ \citep[][]{Christensen08-2008A&A...490...45C} and are 
averaged over the half-light radius of ESO 184--G82 \citep[4 kpc at R-band,][]{Michalowski14-2014A&A...562A..70M}. 
{
Since the depletion time characterises how fast the galactic gas reservoir would be depleted by star formation (at a given SFR),  
it must encompass both the currently star forming material and the non-star forming gas. Therefore, the total (atomic+molecular) gas 
%both the atomic and molecular phases 
should be taken into account when evaluating the galactic-wide star formation regime.
This is especially important in the case of  star-forming dwarfs where   atomic gas forms a significant  fraction of (or even dominates) the total gas mass. 
}
}
%Note that for galaxy-wide scale measurements of depletion time in nearby star-forming galaxies, especially dwarf galaxies, the total (atomic+molecular) gas should be taken into account as atomic gas forms a significant  fraction of (or even dominates) the total gas mass.  
\citet[][]{Lee09-2009ApJ...692.1305L} used the equivalent width of \ha\, emission line ($EW_{\ha}$) and  showed that starburts dwarf galaxies can be robustly classified to have 
$EW_{\ha} >$ 100 \AA. 
The  $EW_{\ha}$ of ESO 184--G82 \citep[56.0 \AA,][]{Kruhler17-2017A&A...602A..85K} thus further excludes the galaxy from being amongst the starburst dwarfs. 
%The definition of starburst dwarf galaxies based on the equivalent width  of \ha\, emission line also places ESO 184--G82 amongst dwarfs without a starburst mode of star formation \citep[see][]{Lee09-2009ApJ...692.1305L}. 
It is therefore clear that the galaxy-scale properties of ESO 184--G82 do not show any evidence whatsoever for the presence of  global 
starburst mode of star-formation in the galaxy.   

{
The \hicm mapping of ESO 184--G82 identified a companion galaxy about 12 kpc from it, and also showed the presence of a high column density gas ring in ESO 184--G82 within which actively star-forming regions and the GRB reside \citep[][]{Arabsalmani15-2015MNRAS.454L..51A, Arabsalmani19-2019MNRAS.485.5411A}.    
Through multi-wavelength observations and simulations,    \citet[][]{Arabsalmani19-2019MNRAS.485.5411A} 
demonstrated  that  a collision between the GRB host   and its companion has given rise to the formation of the  dense gas ring and  has resulted 
in the enhancement of star formation  in the  ring, where the  GRB progenitor was formed. 
%has given rise to the gas and star formation properties of ESO 184--G82 and has resulted  in the enhancement of star formation in the dense gas ring.  
In order  to identify  the properties of molecular gas in this interesting system  and to investigate the ISM conditions in  the vicinity of the GRB, 
%and  to investigate  whether the GRB progenitor was formed in a starburst mode of star formation, 
we present  the  Atacama Large Millimetre Array (ALMA) observations of molecular gas  at $\sim$50 pc scales 
in the metal poor ($Z\sim0.28-0.44\,Z_{\odot}$) host galaxy of  GRB 980425. 
}
We describe  the details of observations and data reduction   in  Section \ref{sec:obs}. The measurements and results are presented 
in Section \ref{sec:res}. A detailed  discussion and the interpretations of results are given in Section \ref{sec:dis}. We summarise our 
findings  in Section \ref{sec:sum}.

%---------------------------------------------------------------------------------------------------
%---------------------------------------------------------------------------------------------------

\section{Observations and data analysis}
\label{sec:obs}
\subsection{ALMA}
We used the Band-6 receivers of the ALMA 12-m Array in C43-5 configuration (with a  maximum baseline of 1400 m) to map the CO(2-1) 
emission of ESO 184--G82  on November 16, 2018 (project code: 2018.1.01750.S; PI: Arabsalmani).  
The maximum recoverable scale for this configuration is 6.7\arcs, allowing the  diffuse CO emission to be recovered  on scales smaller 
than $\sim$1.2 kpc (projected) at the distance of the galaxy \citep[37.3 Mpc using   a flat $\Lambda$CDM with cosmological parameters  from][]{Planck18-2018arXiv180706209P}.  
The observations were for a total time of 5 hours and were conducted with a high-spectral-resolution frequency-division mode spectral window  
centred on 228.534 GHz, covering the redshifted CO(2-1) emission line with a usable bandwidth of  1.875 GHz. This spectral window was sub-divided into 1920 channels, yielding  a velocity resolution of $\sim$1.3 \kms, 
and a total velocity coverage  of  $\sim$2460 \kms. 
The primary beam was  centred on the GRB position {\citep[see][for determination of the GRB position with an error of 0.1\arcs]{Wieringa98-1998GCN....63....1W}. }

\begin{figure*}[]
\centering
\includegraphics[width=0.95 \textwidth]{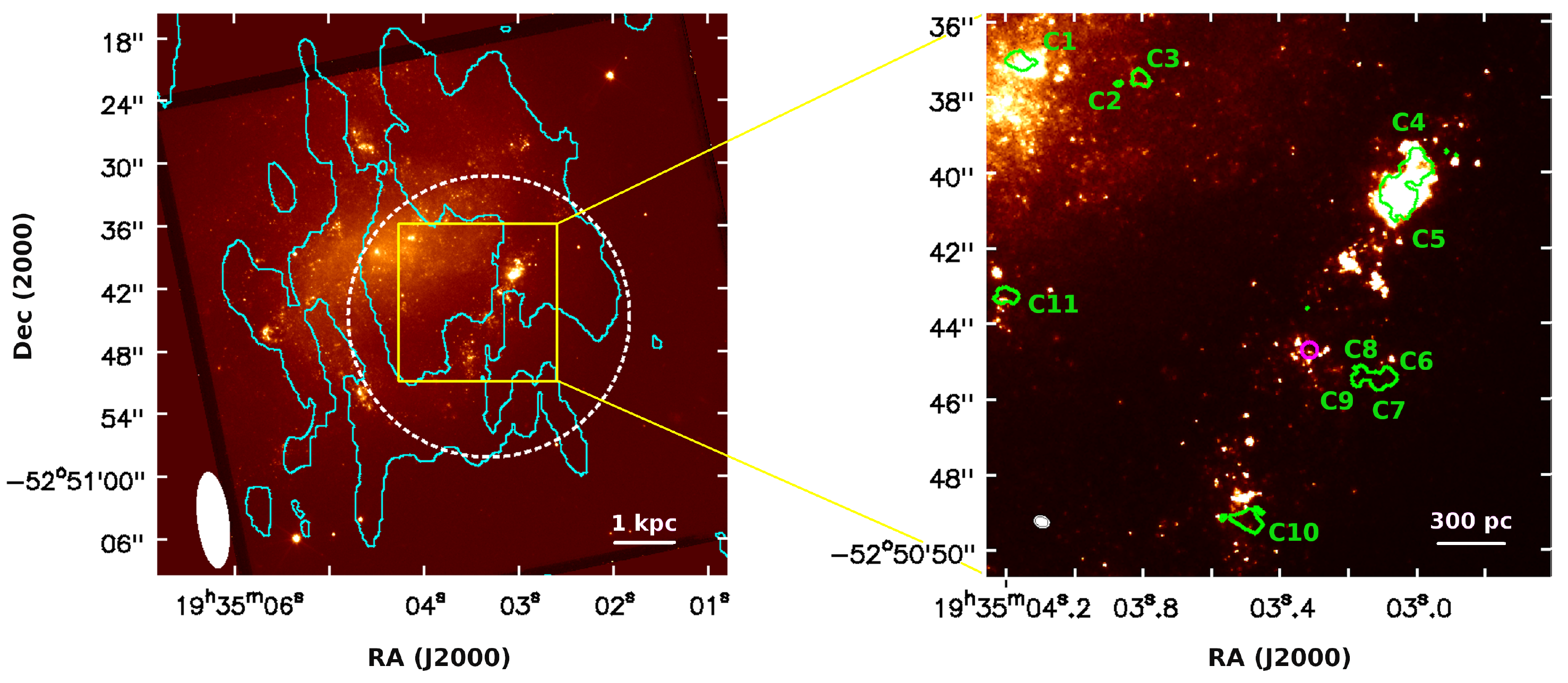}
\caption{\textit{Left:} The HST image of the GRB host galaxy obtained with the MIRVIS/Clear filter, overlayed with the  contour (in cyan) of  \hicm 
 total-intensity  \citep[see][]{Arabsalmani19-2019MNRAS.485.5411A}. The \hi contour is at  a column density of 
$\rm 6.0 \times 10^{20}\, cm^{-2}$ (3$\sigma$ significance) and marks the ring-like structure of high column density atomic gas 
in the galaxy.  
The synthesised beam of the \hicm observations  is shown in the bottom-left corner \citep[for details see][]{Arabsalmani19-2019MNRAS.485.5411A}. 
The dashed  circle shows  the primary beam of ALMA 
in band-6, centred on the GRB position. The yellow box shows the frame of the panel to right. 
\textit{Right:} Contours of CO(2-1) emission at 4$\sigma$ significance obtained from ALMA observations,  overlayed on the HST image.  C1 to C11 mark the 
identified molecular gas clumps. The GRB position is marked with a magenta circle. The synthesised beam of ALMA is shown in the bottom-left corner.     
\label{fig:hst-alma}}
\end{figure*}

The data are calibrated using the ALMA Science Pipeline, part of the  {\sc{Common Astronomy Software Applications (casa)}} 
package \citep[][]{McMullin07-2007ASPC..376..127M}. We then image  the data using the {\sc{casa}} task {\sc{tclean}} 
and  produce  a spectral cube with  a velocity resolution of $\sim$3 \kms. {With  Briggs weighting with robust=0.5}, we get a synthesised 
beam size of 0.35\arcs$\times$0.25\arcs\, (62 pc $\times$ 44 pc at the distance of the galaxy), and  a root-mean-square (rms) 
noise of $\sim$0.4 mJy/beam/channel close to the central frequency of the cube where we expect the line. 
{Due to the primary beam correction applied to the cube,  the rms noise in regions close the edges of the primary beam increases to 
$\sim$0.6 mJy/beam/channel. 
}

We obtain the total-intensity and intensity-weighted velocity field maps of the CO(2-1) emission line using the {\sc MOMNT} task in `classic' {\sc AIPS} \citep[][]{Greisen03-2003ASSL..285..109G}. The {\sc MOMNT} task creates a mask to apply to the spectral cube by smoothing the emission along both the spatial and velocity axes, and then applying a flux threshold to the smoothed emission. This mask is used to choose the pixels in the spectral cube to be integrated for creating the moment maps, thereby ensuring that localized noise peaks are ignored and only true emission that is correlated both spatially and along the velocity axis is selected. 
To create the mask we choose a Gaussian kernel of full width at half-maximum (FWHM) equal to six pixels for smoothing the emission spatially,  Hanning smoothing across blocks of three consecutive velocity channels, and choose a threshold flux of approximately 2 times the root-mean-square (RMS) noise in a line-free channel of the spectral cube.

{
We also make a continuum image combining  the frequency windows of $227.7-228.5$ GHz and $228.7-229.4$ GHz. The continuum emission is detected in only one 
region, centred at RA$=19h35m03.073s$ and Dec$=-52d50m40.72s$, corresponding to  the H{\sc ii} region located about 800 pc to the north-west  of the GRB position 
(see Section \ref{sec:dis} for details). We measured  an integrated flux of $0.197 \pm 0.064$ mJy for the continuum in this region 
using the {\sc{imfit}} task in the {\sc{casa}} package. The measured flux is consistent with the Spectral Energy Distribution presented in 
\citet{Michalowski14-2014A&A...562A..70M}.
}

\subsection{Ancillary data}
%In addition, we use an {Hubble Space Telescope} ({HST})  image of ESO 184--G82  presented in \citet[][]{Fynbo00}. This image was obtained with the MIRVIS/Clear filter centred at 5737.453 \AA\, on June 11, 2000 with a total exposure time of 295 seconds (Program ID: GO-8640, PI: Holland). 
%The data analysis of this {HST} image is described in \citet[][]{Fynbo00}. 
We  use an {Hubble Space Telescope} ({HST})  image of ESO 184--G82, obtained with the MIRVIS/Clear 
filter centred at 5737.453 \AA\, on June 25, 2000 with a total exposure time of 700 seconds (Program ID: GO-8648, PI: Kirshner). 
This image, previously presented in \citet[][]{Kouveliotou04-2004ApJ...608..872K}, was aligned and combined using drizzle, with a 
plate scale of 0.025\arcs\, per pixel (approximately half the native plate scale). The astrometric alignment was 
subsequently refined using the Gaia catalog, {with typical uncertainties of 0.025\arcs.} 
We  also obtain  archival Integral Field Unit (IFU) observations of ESO 184--G82  done using  the MUSE instrument 
on the {Very Large Telescope} ({VLT}) on May 16, 2015 (Program ID: 095.D-0172(A); PI: Kuncarayakti). 
These data have a spatial resolution of $\sim$0.84\arcs\, (equivalent to $\sim$150 pc in projected distance),  
and are already presented in \citet[][]{Kruhler17-2017A&A...602A..85K}. 
{We use the processed MUSE data cubes from the ESO-advanced data products  (Program  ID: 095.D-0172(A)) 
and  apply the astrometric alignment using the Gaia catalog (with typical uncertainties of 0.2\arcs, 
the pixel size in the MUSE cube). 
}
We then make   the total-intensity map of the \ha\, emission line  using the {\sc{immoment}} 
task in the {\sc{casa}} package.

%---------------------------------------------------------------------------------------------------
%---------------------------------------------------------------------------------------------------

\begin{figure*}[]
\centering
\includegraphics[width=1.0 \textwidth]{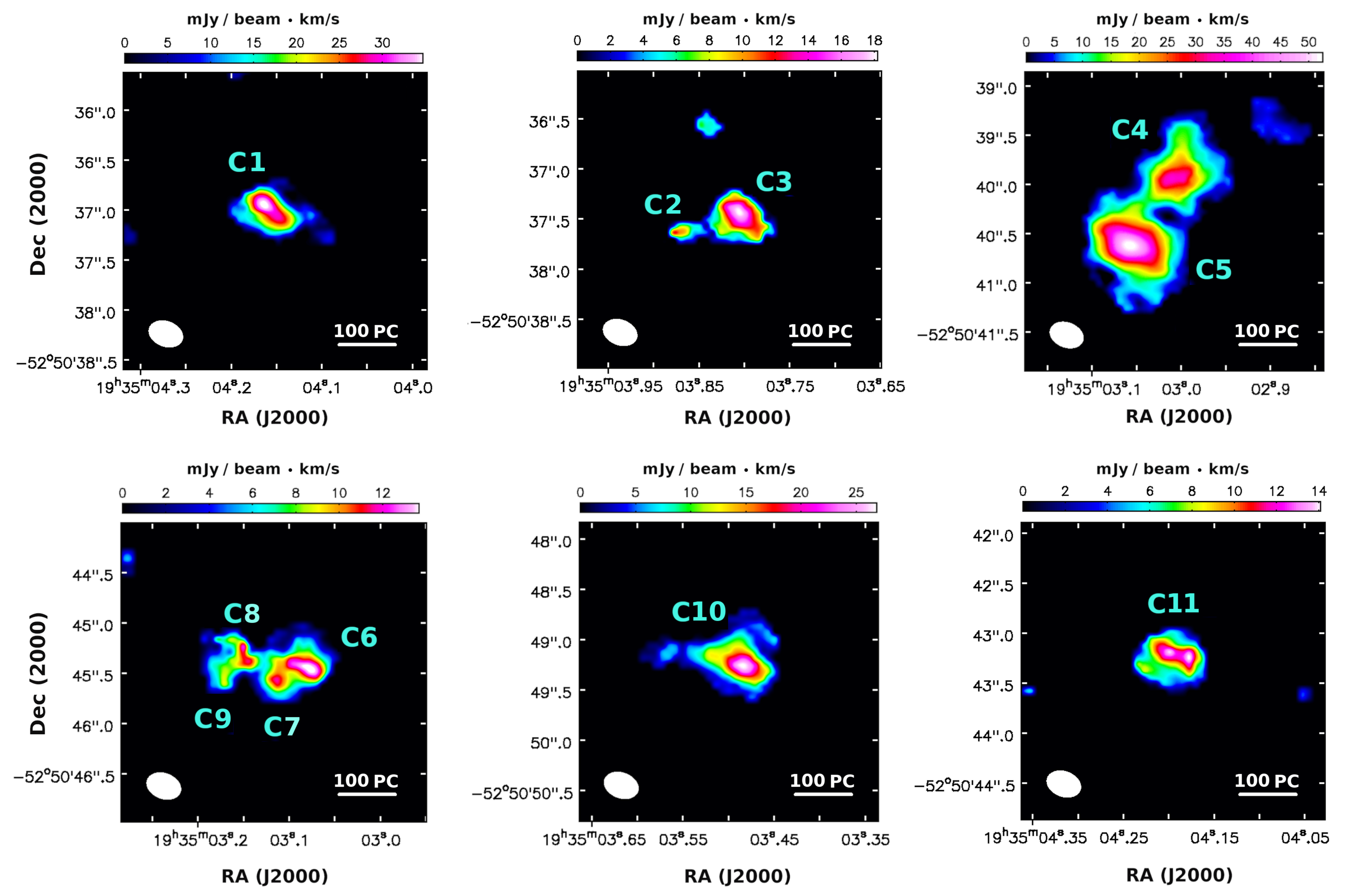}
\caption{The total-intensity maps of CO(2-1) emission line, showing the eleven identified gas clumps  (see Figure \ref{fig:hst-alma} for the location of the clumps in the galaxy). 
The colorbars show  the integrated intensity of CO(2-1) per beam in each clump.  
\label{fig:gmcs}}
\end{figure*}

\section{Measurements and Results}
\label{sec:res}

{The contours of CO(2-1) emission overlayed on the HST image of ESO 184--G82, in the region covered by the 
primary beam of ALMA in band-6, are shown in the right panel of Figure \ref{fig:hst-alma}. These mark the regions  
with significant emission, belonging to  molecular gas complexes. The spatial resolution of the ALMA observations allows 
for the identification of  distinct components in these gas complexes. Given the size of the synthesised beam 
of our ALMA observations (62 pc $\times$ 44 pc), each of these components  can  potentially contain a number of 
distinct sub-structures which we refer to as clumps.  

In order to identify the clumps, we identify the local peaks of emission in the moment-0 (total-intensity) map of 
CO(2-1) emission line that are separated by at least one synthesised beam. These  are the pixels with flux densities  
larger than all the neighbouring  pixels within a  synthesised beam. 
Note that we use the moment-0 map to identify emission peaks and not the cube itself, given that the signal-to-noise 
of the observation is not very high, and the fact that the moment maps were created in such a way (see Section 2.1) that 
localized noise peaks are ignored and only true emission peaks are picked up. 
We identify the area associated with each of the local peaks using the moment-0 map.
For adjacent clumps within the same complex, we use the contours of emission to identify the line of minima between 
the two clumps and separate  the regions accordingly.
We then use the data cube and investigate  the significance of the emission in the  associated areas. If the emission in 
the region is detected in at least two adjacent channels, and with a total significance of at least 5$\sigma$, we identify 
it as a clump. Otherwise we discard the region. 
Furthermore, we look at the  position-velocity diagram of each of the identified clumps and search for sub-structures  
that might be spatially coincident, but kinematically distinguishable. Such a search, however, did  not reveal any such 
structure in any of the gas clumps. 

Using the above mentioned method, we identify a total of eleven molecular gas clumps in the galaxy, marked by C1 to C11 in the right 
panel of Figure \ref{fig:hst-alma}. These clumps are clearly visible in the CO(2-1) total-intensity maps  presented 
in Figure \ref{fig:gmcs}.  
Seven of   these gas clumps (C4-C10) are within the \hi gas ring (see the left panel of Figure \ref{fig:hst-alma}) which was 
identified through  \hicm emission observations presented in \citet[][]{Arabsalmani19-2019MNRAS.485.5411A}. 
The other four gas clumps are located   in the central regions of the  galaxy. 
}

\begin{figure*}[]
\centering
\begin{tabular}{ccc}
\includegraphics[width=0.33 \textwidth]{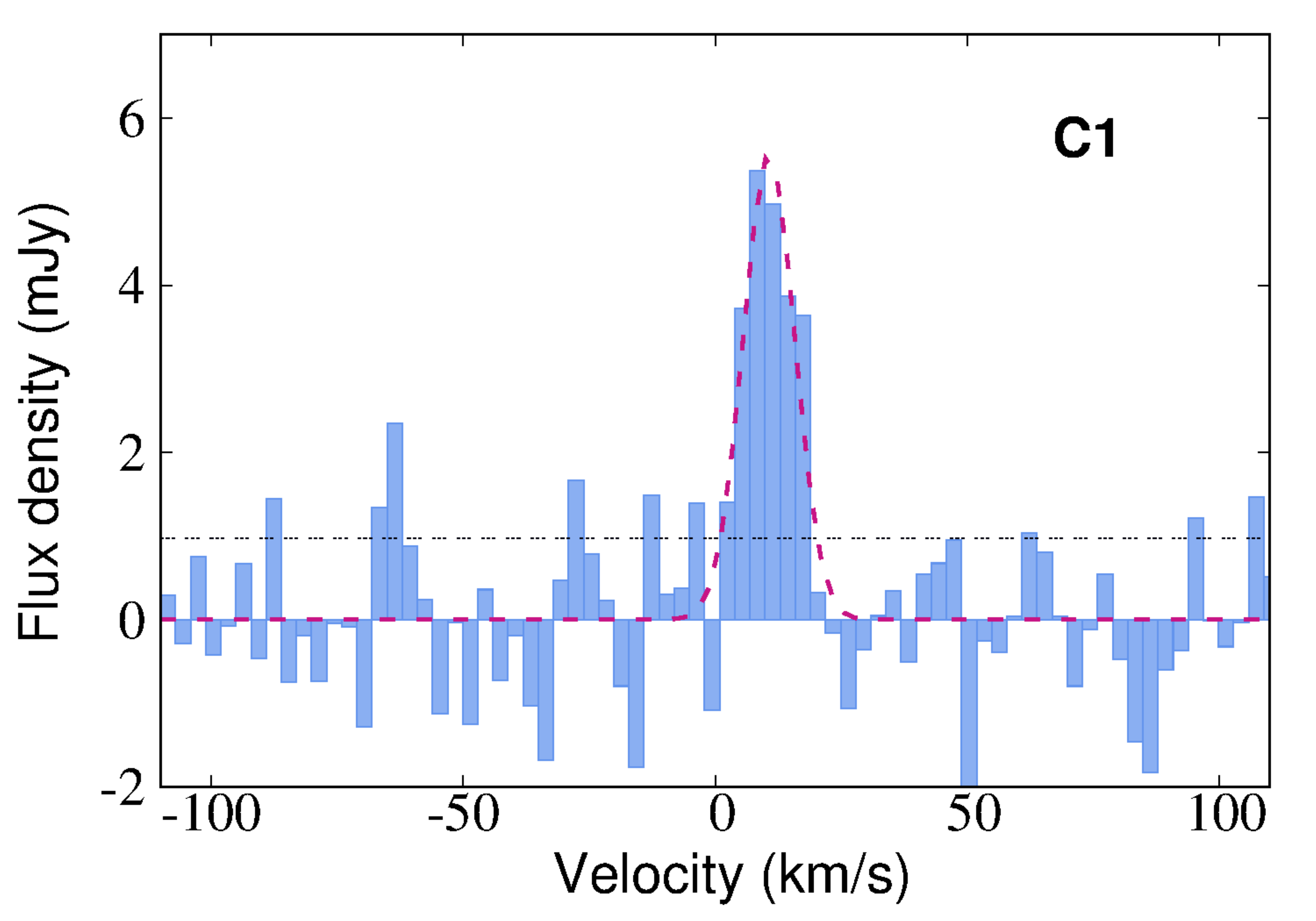}
&
\includegraphics[width=0.33 \textwidth]{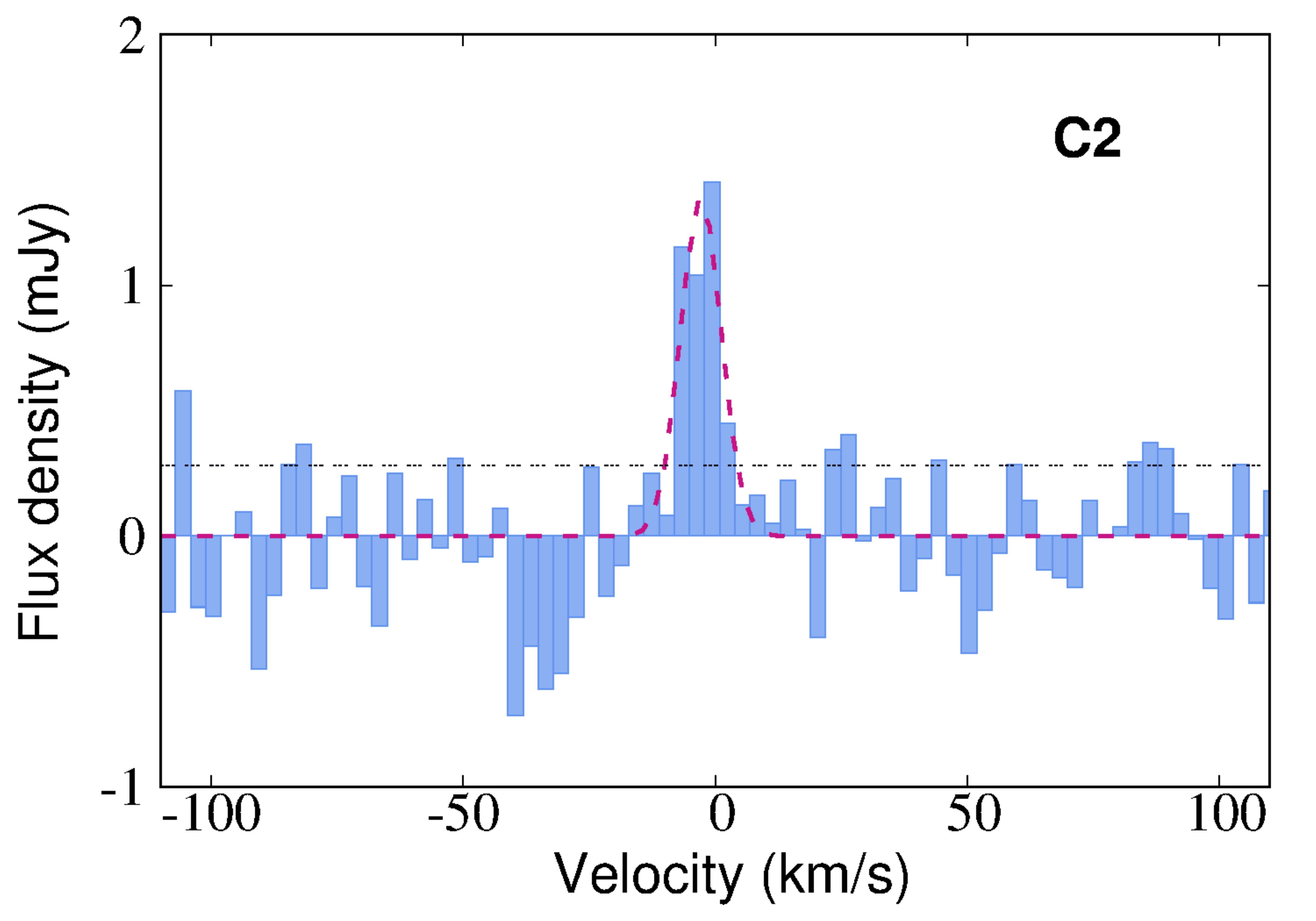}
&
\includegraphics[width=0.33 \textwidth]{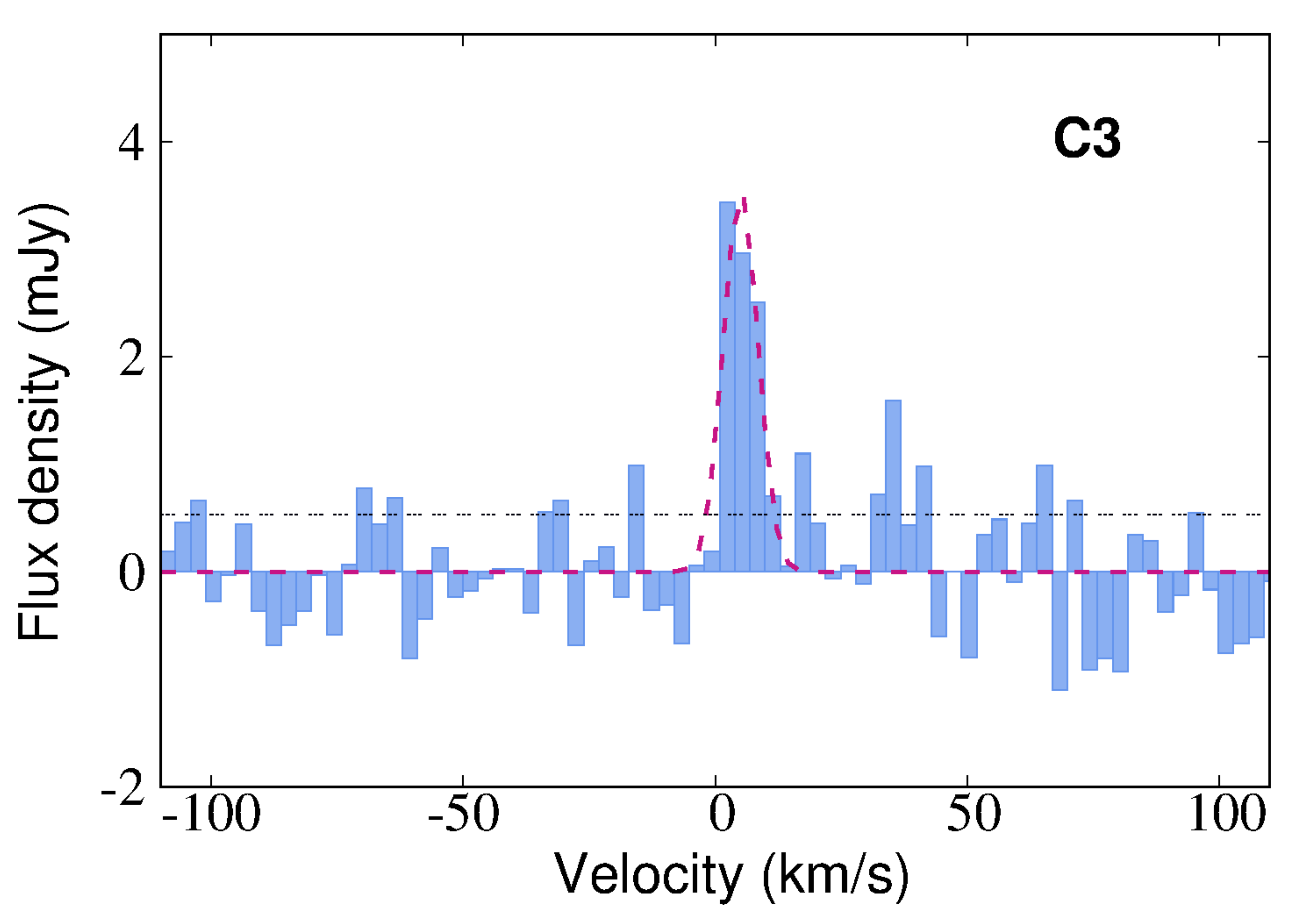}
\cr
\includegraphics[width=0.33 \textwidth]{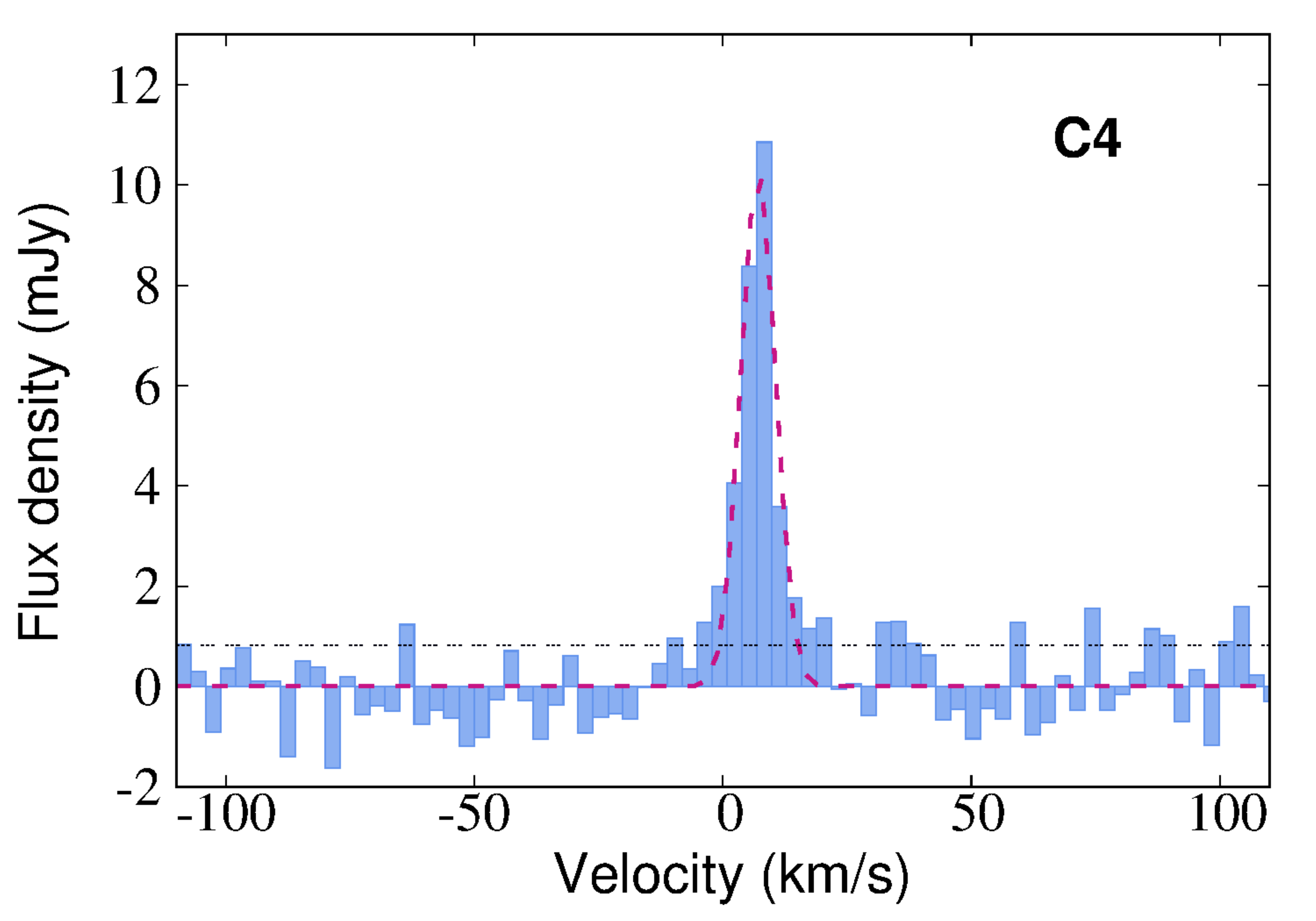}
&
\includegraphics[width=0.33 \textwidth]{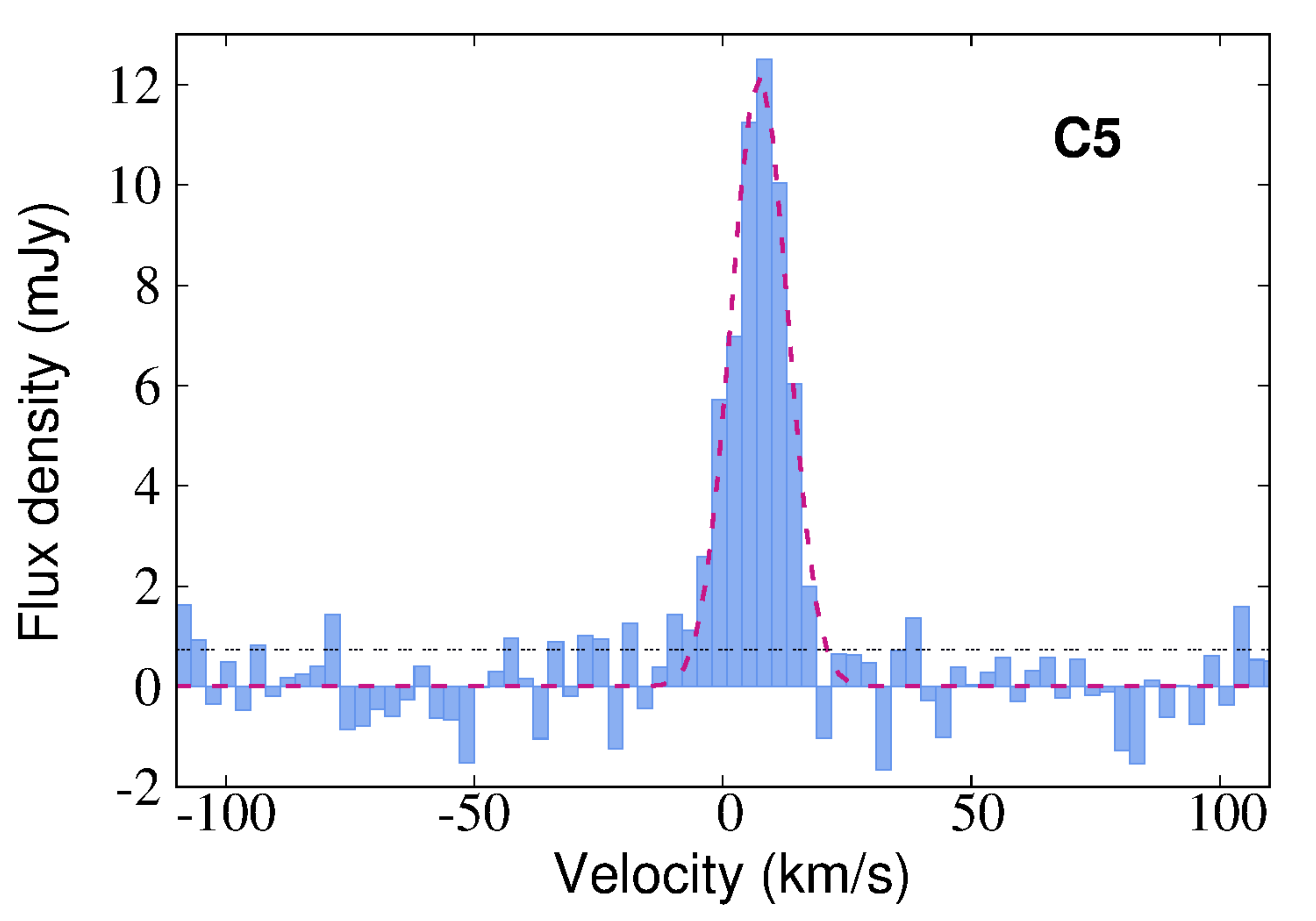}
&
\includegraphics[width=0.33 \textwidth]{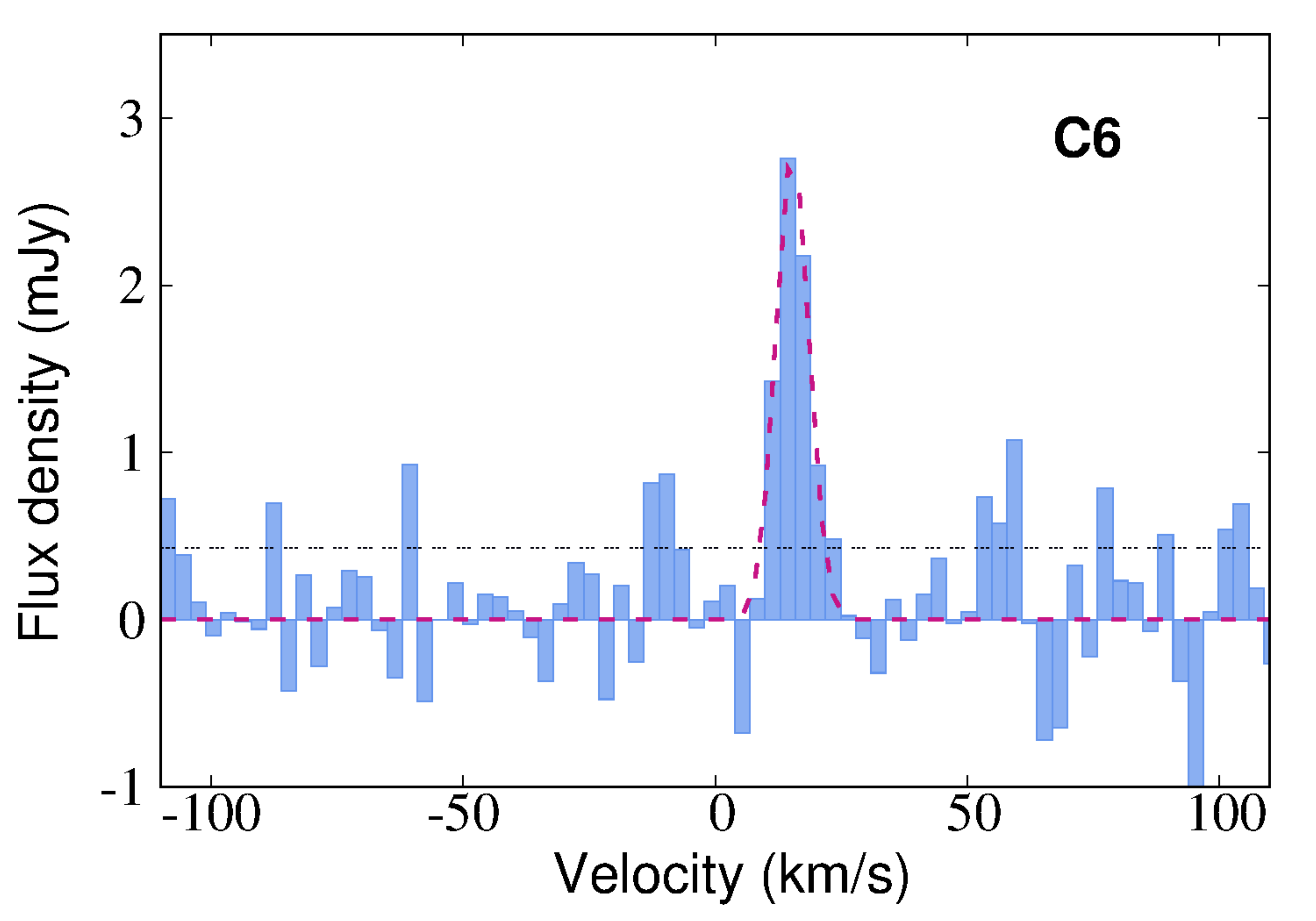} 
\cr
\includegraphics[width=0.33 \textwidth]{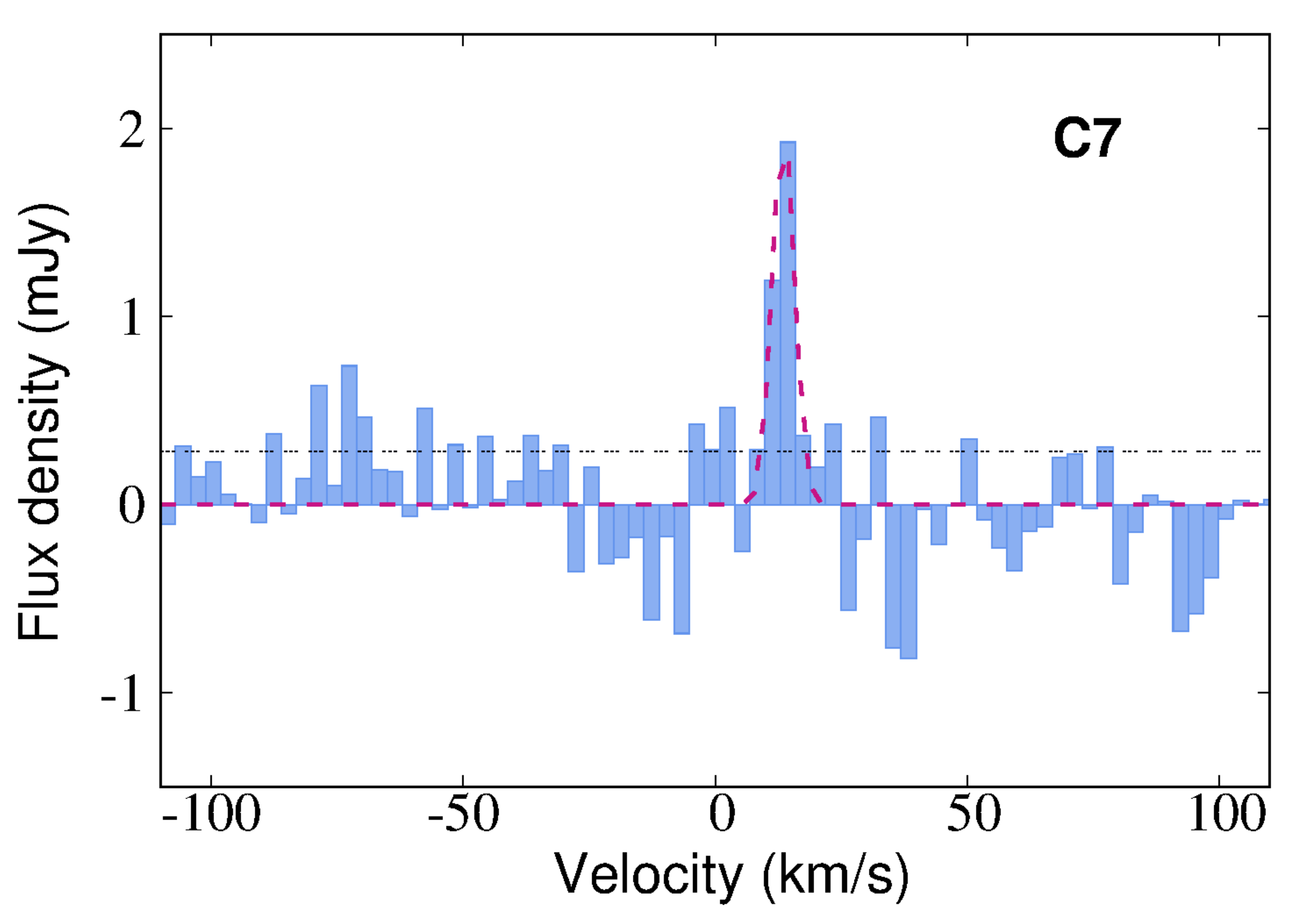}
&
\includegraphics[width=0.33 \textwidth]{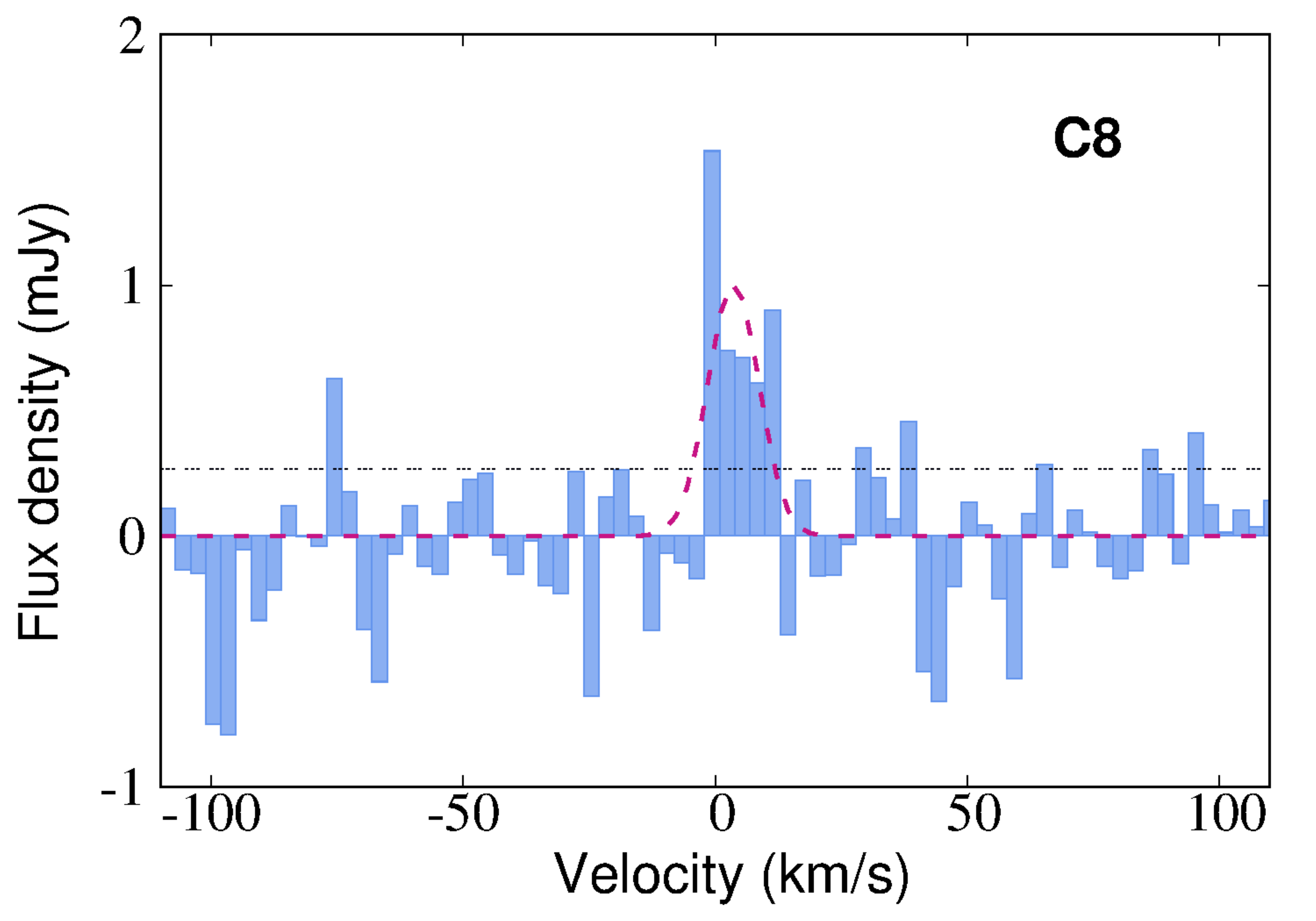}
&
\includegraphics[width=0.33 \textwidth]{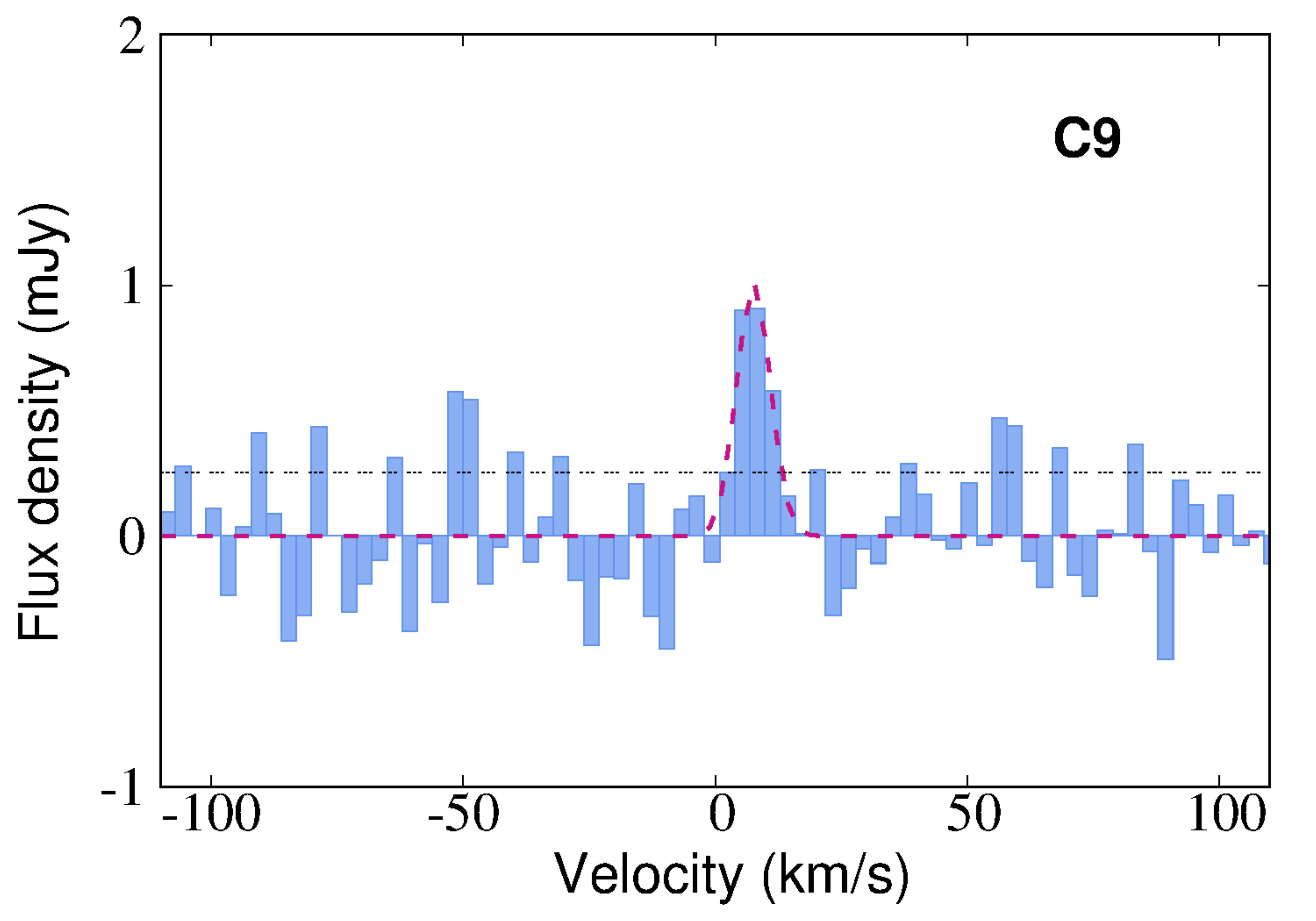}
\cr
\includegraphics[width=0.33 \textwidth]{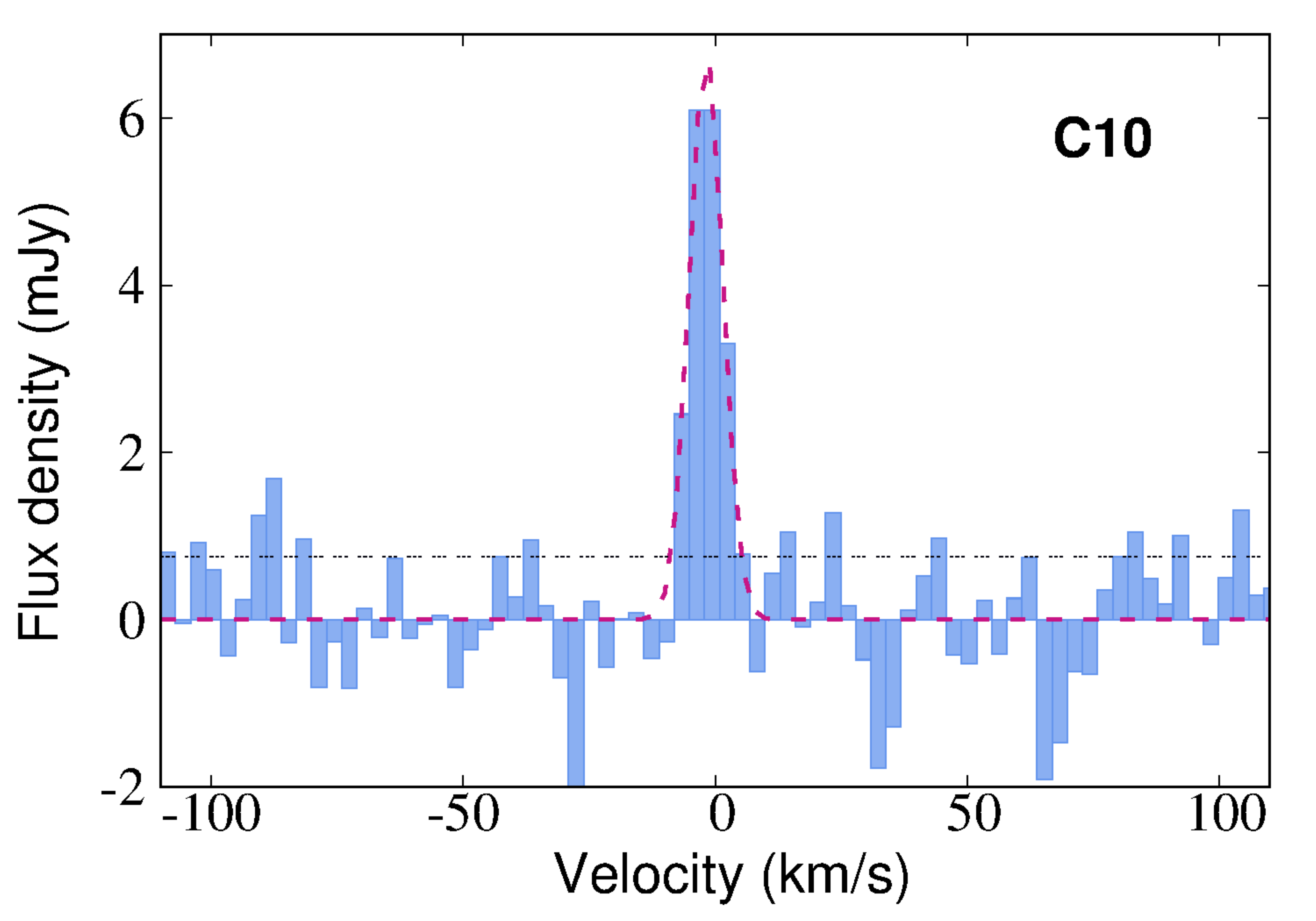}
&
\includegraphics[width=0.33 \textwidth]{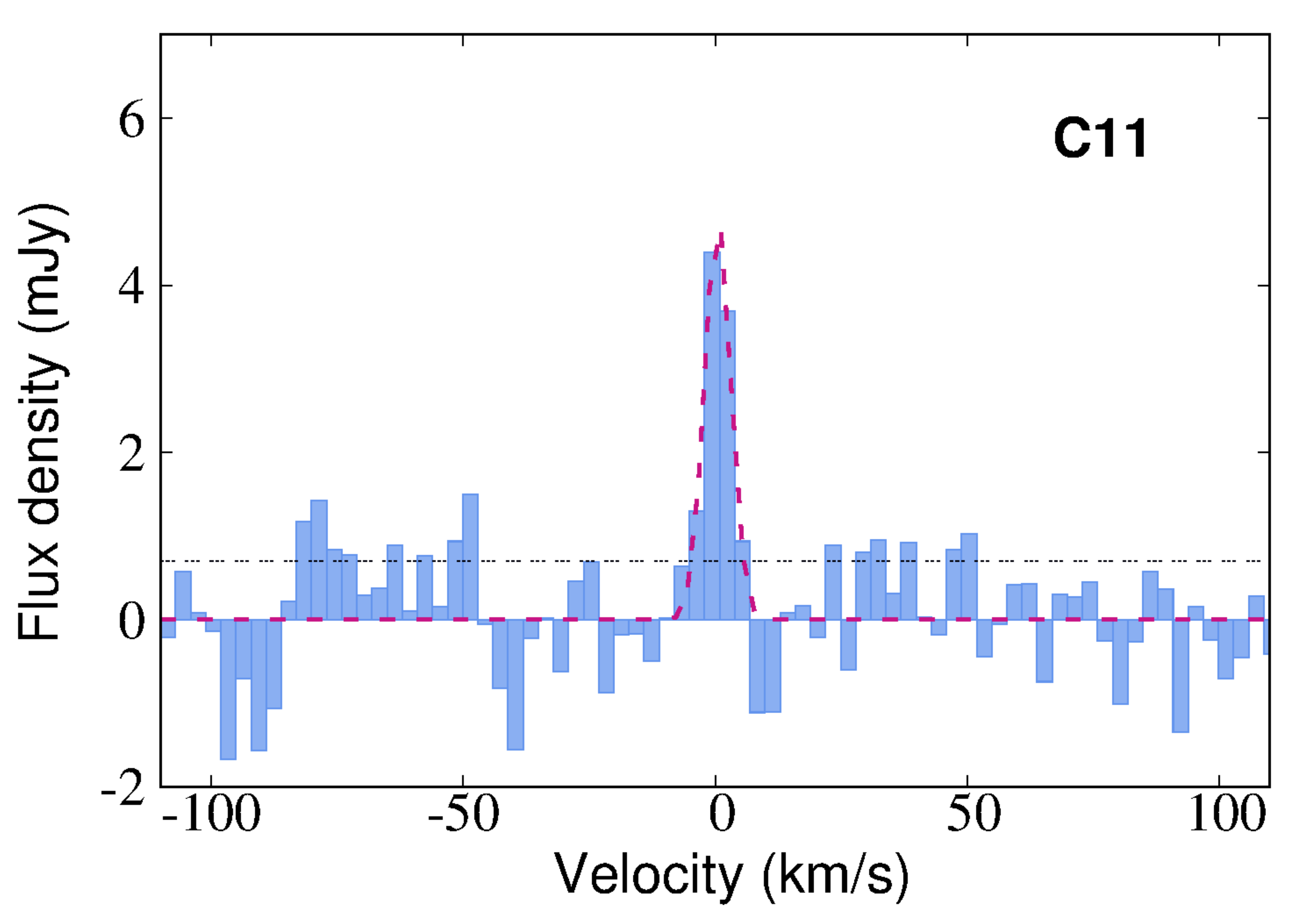}
&
\end{tabular} 
\caption{
The CO(2-1) spectra of the eleven identified  molecular gas clumps in ESO 184--G82. The x-axis 
shows the velocity with respect to $z=0.0086$, the redshift of the galaxy obtained from the \hi spectrum \citep[][]{Arabsalmani15-2015MNRAS.454L..51A}. 
The best-fitted Gaussian functions are shown with red dashed lines. The rms noise level in each spectrum is marked by 
a black dashed line. %Due to the primary-beam response,  the rms values  differ in different regions, with the lowest rms noise level at the centre of the primary beam. 
Note the different scales on the y-axis in different panels.     
\label{fig:spec}}
\end{figure*}

The spectra of CO(2-1) emission line for the  eleven identified molecular clumps  are shown in Figure \ref{fig:spec}. 
We measure the integrated flux density of CO(2-1) emission line ($S\,dv$) by integrating over the adjacent  
channels with fluxes above the rms noise, and obtain the brightness temperature luminosity  of the CO(2-1) 
emission line  ($L^{\rm T}_{\rm CO(2-1)}$) for each of the clumps. 
{
The significance of detection for each clump is calculated by dividing the total flux by $\rm \sqrt{N} \times$ rms-noise-per-channel,  
where $N$ is the number of adjacent  channels with fluxes above the rms noise.  
}
We find that the  clumps in ESO 184--G82 on average have 
higher $L^{\rm T}_{\rm CO(2-1)}$ values   than clumps in nearby compact and starburst dwarfs \citep[see for e.g.,][for 
molecular gas studies of nearby dwarf galaxies based on observations with similar spectral and spatial resolutions 
to our ALMA observations presented in this paper]{Taylor99-1999A&A...349..424T, Walter01-2001AJ....121..727W, 
Leroy06-2006ApJ...643..825L}. 
 
{To measure the velocity dispersions,   we find  the  maximum value in the moment-2 map, $\sigma_v$, for each clump and 
correct it  for the spectral resolution of the cube. We then convert it   to the FWHM velocity, FWHM$_{\rm mom-2}$ \citep[see][]{Rosolowsky06-2006PASP..118..590R}. 
%If the spectrum has a Guassian shape, $\sigma_v$ should be comparable with the standard deviation of the Gaussian. 
We also fit  Gaussian functions to the spectra of all the clumps (see Figure \ref{fig:spec}) and obtain the FWHM of the fitted 
Gaussians, FWHM$_{\rm Gauss}$, also corrected for the spectral resolution of the cube. The FWHM values measured 
from both methods for the clumps are  listed in Table 1. 
We find the two measurements   to be consistent  for all the clumps but C2. This  might indicate  the presence of 
kinematically distinct sub-structures in C2, which we can not distinguish given  the spectral  resolution of 
our observations.  
%In all the analysis we use the velocity dispersion derived from the moment 2 maps.  
%C1  in fact appears to have  the    largest $L^T$ amongst all the molecular clumps in the four galaxies. 
}
%The eleven clumps  in ESO 184--G82 have velocity dispersions of $\sim 5-16$ \kms\ (listed in Table 1), similar to those of clumps  in nearby compact and starburst dwarf galaxies.  

In order to measure the sizes of the clumps, we  fit 2D-Gaussian functions to the moment-0 maps of the clumps 
and correct the measured  standard deviations of the Guassians for the spatial  resolution of the observations by 
deconvolving the synthesised beam from the measured values. We  find the deconvoled values to be comparable to or smaller than 
the major axis (and in some cases the minor axis) of the synthesised beam ($\lesssim 44-62$ pc). The size measurements are therefore 
not reliable and  hence we do not report these measurements. Deeper observations with better spatial resolution are required 
in order to measure the sizes of the molecular clumps in ESO 184--G82.

{To  estimate the molecular gas masses of the clumps we need   to   adopt a  CO-to-molecular-gas 
conversion factor, $\alpha_{\rm CO}$. Extensive studies have shown that $\alpha_{\rm CO}$ increases with decreasing metallicity, 
especially  below $\sim 0.5\,Z_{\odot}$ \citep[see for a review on $\alpha_{\rm CO}$]
[and the references therein]{Bolatto13-2013ARA&A..51..207B}. The values of $\alpha_{\rm CO}$ at low metallicities has been determined 
with a large scatter (more than an order of magnitude) based on both observations and theoretical models  \citep[see Fig.9 in][]
{Bolatto13-2013ARA&A..51..207B}. 
In addition to this, the significant uncertainty on the emission-line metallicity measurements 
due to the choice of calibration \citep[see][and references therein]{Kewlwy19-2019ARA&A..57..511K} makes it even harder to measure 
molecular gas mass   with precision using CO emission. 
\citet[][]{Christensen08-2008A&A...490...45C} and \citet{Kruhler17-2017A&A...602A..85K} used the IFU observations 
of ESO 184--G82 and obtained  the   metallicity maps of  the galaxy. While both studies found the metallicity spread 
within the galaxy to be $\sim 0.3$ dex, they reported different   values  in each region of the galaxy depending on the 
adopted  calibration \citep[see for e.g., the  variation in the metallicity measurements  at the position of GRB in Table 1 of]
[]{Kruhler17-2017A&A...602A..85K}. 
We  therefore  adopt a full metallicity range of $0.28-0.44\,Z_{\odot}$ reported by \citet{Kruhler17-2017A&A...602A..85K} for the average 
metallicity of the galaxy and estimate the molecular gas mass of all the clumps over this metallicity range.  Note that this range also 
covers the measurements provided by \citet{Christensen08-2008A&A...490...45C}. Within this metallicity range, we adopt the metallicity-dependent   $\alpha_{\rm CO}$ suggested by 
\citet[][]{Wolfire10-2010ApJ...716.1191W} which provides one of the better fits to the estimated values of the conversion 
factor from observations \citep[][]{Bolatto13-2013ARA&A..51..207B}. We therefore use a range of  $\alpha_{\rm CO} = 12-40\, M_{\odot}\,\rm(K\,km\,s^{-1}\,pc^{2})^{-1}$ 
for estimating the molecular gas masses of all the clumps in our analysis.   
We also assume a  ratio of 0.8 for the brightness temperature luminosity  of the CO(2-1) to CO(1-0) lines, which is the  
typical value derived for nearby star-forming galaxies \citep[][]{Leroy09-2009AJ....137.4670L}. 
%The uncertainty on the ratioof the luminosity temperatures of the CO(2-1) to CO(1-0) lines is much less than the expected uncertainty  on the conversion factor. 
Note that  the uncertainty  on the ratio of the luminosity temperatures of the CO(2-1) to CO(1-0) lines is expected to be 
small in comparison to that on the $\alpha_{\rm CO}$ conversion factor. This ratio is also shown  not to be much sensitive to CO excitation 
conditions \citep[see][and references therein]{Daddi15-2015A&A...577A..46D}. The measured properties of the eleven clumps are summarised in Table \ref{tab:gmcs}.
}

\begin{table*}
\centering
%\begin{threeparttable}
\caption{Properties  of  molecular gas clumps in ESO 184--G82.}
\label{tab:gmcs}
\begin{tabular}{ccccccccc}
\toprule
Clump   &  S/N & $S\,dv$   &   $L^{\rm T}_{\rm CO(2-1)}$    & FWHM$_{\rm mom-2}$  &  FWHM$_{\rm Gauss}$   &   $M_{\rm mol}$   & SFR & $\tau_{\rm dep}$ \\
        &      & (mJy\,\kms)      &   (K\,\kms\,pc$^2$)        & (\kms)     & (\kms)  	&  ($10^5\,M_{\odot}$) & ($10^{-3}\,M_{\odot}\,\rm yr^{-1}$) & (Gyr) \\
\midrule
C1      &  10  & 68.4$\pm$7.1     &    5.7 $\times 10^4$       & 11      &  12  	&   9$-$28     & 1.4 & 0.6$-$2.0\\
\hline
C2      &  7   & 12.1$\pm$1.7     &    1.0 $\times 10^4$       & $<$3    &  9       	&   2$-$5      & \multirow{2}{*}{0.1} & \multirow{2}{*}{5.8$-$19.3} \\
C3      &  9   & 28.6$\pm$3.2     &    2.4 $\times 10^4$       & 6       &  8      	&   4$-$12     &  & \\
\hline
C4      &  14  & 102.5$\pm$7.3    &    8.5 $\times 10^4$       & 11      &  8    	&   13$-$43    & \multirow{2}{*}{45.1} & \multirow{2}{*}{0.08$-$0.26}	\\
C5      &  26  & 177.6$\pm$6.9    &   14.7 $\times 10^4$       & 16      &  13      	&   22$-$74    & &     \\
\hline
C6      &  8   & 23.1$\pm$2.8     &    1.9 $\times 10^4$       & 7	 &  7    	&   3$-$10      &   \multirow{3}{*}{0.36} &   \multirow{3}{*}{2.1$-$7.0}\\
C7      &  7   & 11.2$\pm$1.7     &    0.9 $\times 10^4$       & 5	 &  4    	&   1$-$5      & & \\
C8      &  8   & 13.4$\pm$1.8     &    1.1 $\times 10^4$       & 13      &  12       	&   2$-$6      & & \\
C9      &  5   &  7.9$\pm$1.5     &    0.7 $\times 10^4$       & 7       &  7      	&   1$-$3      & & \\
\hline
C10     &  11  & 55.8$\pm$5.0     &    4.6 $\times 10^4$      & 9	 &  9    	&   7$-$23     &  0.4 & 1.8$-$5.9\\
\hline
C11     &  7   & 30.7$\pm$4.1     &    2.6 $\times 10^4$      & 4	 &  5    	&   4$-$13     &  0.2 & 1.8$-$5.8\\
\bottomrule
\end{tabular}
\flushleft
{Columns: (1) clump identifier, (2) significance of the detection, (3) integrated flux density of CO(2-1) 
emission line, (4) brightness temperature luminosity of CO(2-1) emission line, (5,6) the velocity dispersion measured from the 
moment-2 maps and Gaussian fitting, respectively,  corrected for the spectral resolution, (7) estimated molecular gas mass for 
the molecular gas clump, assuming a range of $\alpha_{\rm CO}=12-40\,M_{\odot}\,\rm(K\,km\,s^{-1}\,pc^{2})^{-1}$, (8) 
star formation rate of the gas complex, and (9) molecular gas depletion time of the gas complex.  
}  
%\end{threeparttable}
\end{table*}

\begin{figure*}[]
\centering
\includegraphics[width=0.95 \textwidth]{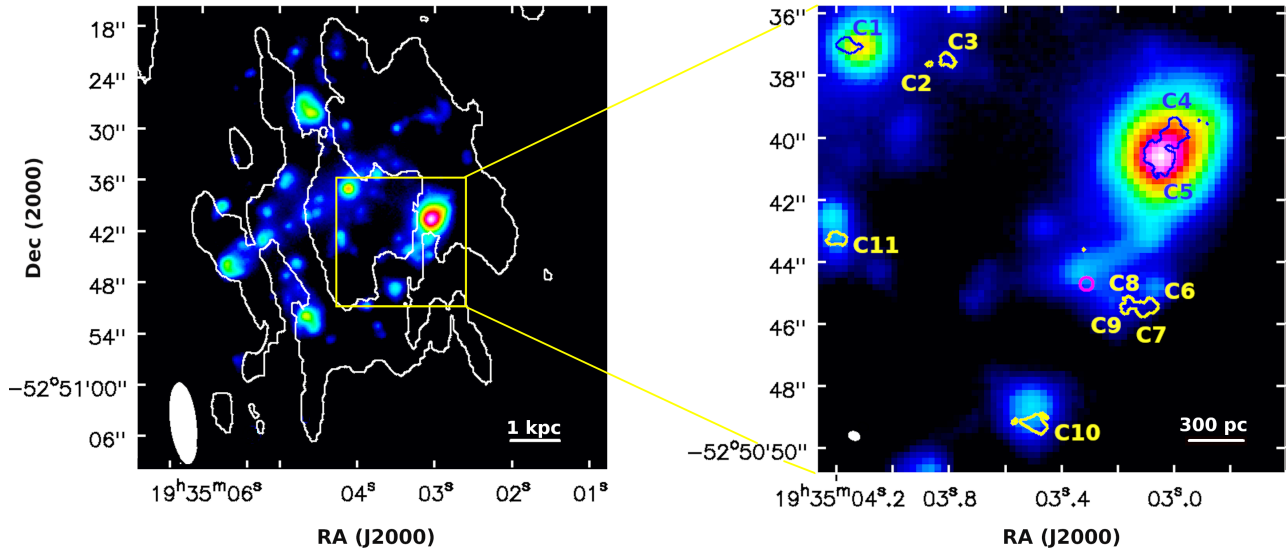}
\caption{\textit{Left:} The total-intensity map of the  \ha\, emission line (made using the MUSE cube) from ESO 184--G82  overlayed with the  contour (in white) of  \hicm 
 total-intensity  \citep[see][]{Arabsalmani19-2019MNRAS.485.5411A}. The \hi contour is at  a column density of 
$\rm 6.0 \times 10^{20}\, cm^{-2}$ (3$\sigma$ significance) and marks the ring-like structure of high column density atomic gas 
in the galaxy.  
The synthesised beam of the \hicm observations  is shown in the bottom-left corner \citep[for details see][]{Arabsalmani19-2019MNRAS.485.5411A}. 
The yellow box shows the frame of the panel to right. 
\textit{Right:} The total-intensity map of the  \ha\, emission line (made using the MUSE cube) in the same frame as the right panel of 
Figure \ref{fig:hst-alma},  overlaid with the contours of CO(2-1) emission at 4$\sigma$ significance.   C1 to C11 mark the 
identified molecular gas clumps. The GRB position is marked with a magenta circle. The synthesised beam of ALMA is shown in the bottom-left corner.      
\label{fig:muse-alma}}
\end{figure*}

We are unable to determine the surface densities of the clumps due to the insufficient spatial resolution of our 
ALMA observations. But %in order to check whether the clumps locally have starburst mode of star formation, 
we use the MUSE observations to measure the SFR and compute  the molecular gas depletion time ($\tau_{\rm dep}$, 
defined as the ratio of molecular gas mass to the SFR) for the gas clumps. 
The right panel of Figure \ref{fig:muse-alma} shows the total-intensity map of the \ha\, emission line in the same 
frame as the right panel of Figure \ref{fig:hst-alma}. The CO contours overlayed on the \ha\, map mark the locations 
of the molecular gas clumps with respect to the star-forming regions. 
Given the coarser spatial resolution of MUSE observations ($\sim$0.85\arcs\, compared to $\sim$0.3\arcs\, resolution of 
the ALMA observations), we obtain SFR and $\tau_{\rm dep}$ for each of the gas complexes  (instead of measuring it for 
the eleven individual clumps). 
{
This ensures that the SFR measurements are done within regions which have comparable or larger sizes 
with respect to the  spatial resolution of the MUSE cube ($\sim$150 pc). 
The measurements are done in regions which are identified by the CO emission and irrespective of the locations of the \ha\, peaks.    
We calculate the total flux in all the pixels of the \ha\, map  within the  identified region for each of the  gas complexes 
and compute the SFR associated with  the gas complexes using  the calibration of \citet[][]{Kennicutt98-1998ApJ...498..541K}, 
assuming a Salpeter initial mass function (IMF). 
With a galaxy-averaged attenuation of $E_{B-V}$ of 0.05 mag  for ESO 184--G82 and a similar foreground Galactic extinction 
\citep[][]{Kruhler17-2017A&A...602A..85K}, the SFR correction for dust extinction will be insignificant,  
and hence we do not apply such a correction. 
The SFR and  $\tau_{\rm dep}$ measurements for the gas complexes are presented in columns 8 and 9 of Table. \ref{tab:gmcs}.

In any galaxy, the spatial offset between  gas and stars on cloud scales, and also the difference between the gas that  
has fuelled the formation of the stars with the gas coexisting with the stars at the time of observations, imply that it is not possible to  \emph{consistently} measure the 
efficiency of star formation, or the depletion time, on cloud scales. For instance, the depletion time measured at cloud scales  are necessarily 
different from those at galactic scale which account for the entire gas reservoir \citep[see e.g.,][]{Schruba10-2010ApJ...722.1699S}. 
This is because gas depletion time measurements on cloud scales do not reflect the diversity of timescales of the processes involved in triggering 
and regulating star formation. 
Such measurements however can provide important clues regarding the nature of star formation in a galaxy.
For example, spatial and temporal offsets between gas and stars on cloud scales are particularly pronounced in rapidly evolving systems 
which  significantly impact the measured gas depletion times on cloud scales \citep[see][in the 
context of interacting galaxies]{Renaud19-2019A&A...625A..65R}.  Without making assertions based on the measured values, we  compare the 
scatter in measured  depletion times of gas clumps in ESO 184--G82 with other studies done on similar scales in other galaxies using a similar approach. 
}

%, and are therefore lower limits to the overall SFRs in the regions as will be measured at larger spatial scales sensitive to both gas and SFR peaks \citep[see][]{Schruba10-2010ApJ...722.1699S, Kruijssen14-2014MNRAS.439.3239K}.This implies that the molecular gas depletion times that we infer are actually longer  than the average  depletion times that will be measured over larger spatial scales in the galaxy.

We find the molecular gas depletion times  to show  a variation of  1.9 dex  (given a fixed value for the $\alpha_{\rm CO}$, as is usually 
assumed for star-forming galaxies in the literature).  
{
This is significantly larger  than the typical spread on  $\tau_{\rm dep}$ at similar spatial  scales  observed in normal star-forming galaxies 
in the nearby Universe. 
\citet[][]{Schruba10-2010ApJ...722.1699S}  measured the $\tau_{\rm dep}$  on 75 pc to 1.2 kpc  scales in M33 
by centering on molecular gas peaks (similar to what we do here), and found it to vary by $<$0.3 dex at each resolution. 
\citet[][]{Bolatto11-2011ApJ...741...12B} found a characteristic molecular gas depletion time (of $\sim$1.6 Gyr) on scales 
of 200 pc in the Small Magellanic Cloud and found it to remain unchanged up to 1 kpc scales. 
The studies of samples of nearby spiral galaxies by \citet[][]{Bigiel11-2011IAUS..270..327B} and \citet[][]{Leroy13-2013AJ....146...19L} 
showed typical spread of 0.5 dex at sub-kpc and kpc scales.  
}
{
The large spread  of 1.9 dex in  the measured  $\tau_{\rm dep}$ values  of the molecular gas complexes in ESO 184-G82 is   comparable with 
the large variations of $\tau_{\rm dep}$ found in  starburst galaxies which show large spatial variations in the physical conditions  
 that trigger  star formation, and where normal modes of star formation co-exist  with highly enhanced regimes 
\citep[see for e.g.,][]{Pereira-Santaella16-2016A&A...587A..44P, Tomicic18-2018ApJ...869L..38T, Renaud19-2019A&A...625A..65R}. 
}

%------------------------------------------------------

\section{Discussion}
\label{sec:dis}

While the global properties of ESO 184--G82 has no indication of a starburst (recall Section \ref{sec:int}),  the $\sim$ 100 pc scale studies of gas and star formation 
in the galaxy suggest otherwise. 
%There are several lines of evidence, indicating the formation of GRB 980425 progenitor  in starburst mode of star formation in ESO 184--G82.  
The large spread  of 1.9 dex in  the measured  $\tau_{\rm dep}$ values  of the molecular gas complexes in ESO 184-G82  demonstrates  the presence of \emph{local} 
starburst modes of star formation in ESO 184--G82, which is  the favourable condition  for the formation of  dense and massive star clusters.    
Indeed, the estimated  molecular gas masses of $\geq 10^{5}\,M_{\odot}$ for the identified  clumps in ESO 184--G82 
are sufficient for  the formation of  massive star clusters with masses larger than $10^{4}\, M_{\odot}$ \citep[considering typical  efficiencies  of 10\% for conversion of molecular gas into stars in  star-forming  galaxies,][]{Ochsendorf17-2017ApJ...841..109O, Grisdale19-2019MNRAS.486.5482G}.  
%The absence of molecular gas at the position of GRB   indicates that molecular  clouds in that region have been dispersed, most likely by stellar feedback. This is consistent  with the age estimate of stars in this region of 5 Myr that \citet[][]{Kruhler17-2017A&A...602A..85K} obtained from the equivalent of \ha\, emission line.  The 3$\sigma$ upper limit  on the  molecular gas mass  in this region  implies  that the molecular gas depletion time is shorter than  0.8 Gyr, compatible with the   starburst mode of star formation in the region.  

{
By assuming $\alpha_{\rm CO} = 12-40\, M_{\odot}\,\rm(K\,km\,s^{-1}\,pc^{2})^{-1}$, we estimate an upper limit of $0.37-1.25$ Gyr for  the  
molecular gas depletion time at the GRB position. 
The upper limit is based on the CO(2-1) emission having a significance of 3$\sigma$ in each channel, and spread over at least two channels (same as 
the detection criteria described in Section \ref{sec:res}), measured  in a region with a projected size of $\sim$ 150 pc (equivalent to 
the spatial resolution of the MUSE observations) and centred at the GRB position. }  
%The 3$\sigma$ upper limit  on the  molecular gas mass  at the GRB position (assuming that the emission is spread over at least 2 channels / 6 \kms) implies  
%\citet[][]{Schruba10-2010ApJ...722.1699S}'s measurements of $\tau_{\rm dep}$ in M33 when centering on molecular gas peaks showed that the depletiontime increases with decreasing spatial scale size \citep[see][for an explanation of the reason behind this bias]{Kruijssen14-2014MNRAS.439.3239K}. Comparing with their results, our measurements of $\tau_{\rm dep}$ over $\sim$150 pc scales is likely to be an upper limit to the `true' depletion time at the GRB position.
{
The estimated upper limit on $\tau_{\rm dep}$ at the GRB position is the shortest depletion time after the WR region  in ESO 184--G82.  
This, combined with the large scatter on $\tau_{\rm dep}$ on cloud  scales in the host 
galaxy (see Section \ref{sec:res}),  strongly suggest that the progenitor of GRB 980425 was formed in an extreme  mode of star formation, like a local starburst. 
}
Note that there are several lines of evidence  for a top heavy IMF  in  starburst regions \citep[e.g.,][]{Schneider17-2018Sci...359...69S, Zhang18-2018Natur.558..260Z}, suggesting  that such regions are ideal birth-places for GRB progenitors.

In \citet[][]{Arabsalmani19-2019ApJ...882...31A} we performed $\sim$300 pc scales studies  of molecular gas in the host galaxy of 
a Superluminous Supernova (SLSN). These are another class of extremely bright explosions of 
massive stars and are found in similar environments as GRBs. Our ALMA  observations of the massive and metal-rich host galaxy of 
SLSN PTF10ptz showed   that the SLSN progenitor was also  formed in starburst mode of star formation 
triggered by  the internal dynamics of the bar in the host galaxy. 
But unlike the case of SLSN PTF10ptz where we found the SN position to be located within  high surface density molecular gas, we do not detect any molecular gas at the position of GRB 980425.

The location of GRB 980425 is offset by about 1.6\arcs\, ($\sim$280 pc) from  the centre  of  the closest clumps, C8-C9 
(see the right panel of Figure \ref{fig:hst-alma}).  This offset is significantly larger than the 
reported systematic offset of $\sim$0.25\arcs\,  between the ALMA and HST positions \citep[see e.g.,][]{Barro16-2016ApJ...827L..32B, Dunlop17-2017MNRAS.466..861D}. 
%One may want to explore  the possibility of an   association between the GRB progenitor and these clumps. 
%Observations of nearby galaxies as well as simulations show that over time young star clusters can get separate  from the gas clouds in which they were formed \citep[][]{Renaud13-2013MNRAS.436.1836R}. This is due to the difference in velocities of gas and stellar components  (typically a few  \kms) and is known as asymmetric drift (of gas with  respect to stars).  
%Considering that the GRB  progenitor is expected to have formed a few Myr ago \citep[with an estimated  mass of $40\,M_{\odot}$, derived from the observed properties of the associated SN;  see][]{Mazzali01-2001ApJ...559.1047M}, the separation between the gas clumps and the GRB position predicts a velocity difference of about 100 \kms\, between stellar and gas components. This  is  much larger that the observed velocity differences between gas clouds  and the young star clusters   in  nearby galaxies, making  the association between C7 and the stellar component in which the GRB is located unlikely. 
It is plausible   that the  GRB progenitor was a runaway star, dynamically ejected  from a cluster formed within this clump. 
Using  N-body simulations \citet[][]{Oh16-2016A&A...590A.107O} showed that young star clusters tend to 
eject their massive stars with velocities that can be larger than one hundred  \kms\, and thus compatible with the 
observed offset. They found the efficiency  of ejecting massive stars to be higher for clusters with higher 
densities, and  to increase with an increase in the mass of the ejected star. This suggests that the massive 
progenitor stars of GRBs can be runaway stars, ejected  from the most dense clusters.  
It is notable that  a significant fraction of the ejected  stars in the study of \citet[][]{Oh16-2016A&A...590A.107O} 
are in  multiple systems. This is compatible with the formation of GRBs  through the interaction of massive 
stars in binary or multiple systems \citep[][]{Chrimes20-2020MNRAS.491.3479C}. 
 
The progenitor of GRB 980425 has an estimated  mass of $40\,M_{\odot}$ derived from the observed properties of the associated SN 
\citep[][]{Mazzali01-2001ApJ...559.1047M}. It therefore   is expected to have formed a few Myrs ago. 
The velocity  required   by  a runaway star  to travel the distance between C8-C9 and the GRB position in 3 Myr is 
$\sim$90 \kms, which   is on the higher side of runaway velocities \citep[see Figure 4. in ][]{Oh16-2016A&A...590A.107O}.  
\citet{Hammer06-2006A&A...454..103H} suggested  that the GRB progenitor is a runaway star that was ejected from the  H{\sc ii} region  
located about 800 pc to the north-west  of the GRB position. This extremely bright H{\sc ii} region has been shown to contain a large number 
of  Wolf-Rayet (WR) stars with ages less than a few Myrs \citep[][]{Hammer06-2006A&A...454..103H, Kruhler17-2017A&A...602A..85K}   
and is referred to as  the WR region in ESO 184--G82. 
C5, the   brightest clump in CO(2-1) emission, is clearly associated with this region. 
Our  2D Gaussian fittings to  the CO and \ha\, intensity maps as well as  the HST image 
show that  the peak of the CO emission in C5 is  coincident  with those  of the stellar component and the \ha\, emission in the WR region. 
This region not only has the  brightest CO(2-1) emission, but also demonstrates  the shortest depletion time.   
The presence of large amount of molecular gas  shows  that  the stellar feedback in the WR region  has not yet dispersed the 
molecular gas. This, together with the short depletion time of molecular gas, points to the   high densities of molecular gas,  as well as 
the presence of a local starburst mode of star formation in the region. 
{
This is in agreement  with the findings of   \citet[][]{LeFloch12-2012ApJ...746....7L} where  an extensive study of the WR region  
based on  infrared observations reveals the starburst nature of the region. %, with a sSFR of $\gtrsim 20$ Gyr$^{-1}$. 
}  
\citet{Hammer06-2006A&A...454..103H} argued that  the progenitor of GRB 980425 was  ejected from 
the WR region (C5), with a  runaway velocity  of $\sim$300 \kms, as   required for the progenitor to travel from 
the WR region to the GRB location in 3 Myr. Such a large runaway velocity, though possible, is very rare \citep[see Figure 4. in ][]{Oh16-2016A&A...590A.107O}. 
Therefore, in a runaway progenitor scenario,  the ejection of the progenitor star(s) from a cluster  formed within C8-C9 
seems more likely than the progenitor being ejected from the WR region.

The position of the GRB coincides with a concentration of stellar emission, as seen on the HST image (Figure 
\ref{fig:hst-alma}), and noted in previous studies. \citet[][]{Kruhler17-2017A&A...602A..85K} estimated the age of young stars in this 
region to be $\sim$5 Myr based on the equivalent width  of \ha\, emission line. 
This makes the  coincidence of a runaway star with the  stellar structure (of the same age) at GRB position very unlikely.   
It is therefore more likely that the  progenitor of GRB 980425  was associated with this stellar structure and that it   formed in situ 
\citep[see also the discussion in][]{Kruhler17-2017A&A...602A..85K}.
%The position of the GRB coincides with a concentration of stellar emission, as seen on the HST image (Figure \ref{fig:hst-alma}), and noted in previous studies. This  suggests that the GRB  is likely to be associated with this stellar structure and that its progenitor  formed in situ. 
The non-detection of molecular gas in this region  indicates that the native cloud of the GRB and the nearby stars has already 
been dispersed, most likely by stellar feedback. This is compatible with the age estimate of stars in this region 
of 5 Myr, i.e. longer than the onset of first type II SNe that ionised the gas.

Using the multi-wavelength observations  and simulations, \citet[][]{Arabsalmani19-2019MNRAS.485.5411A}  demonstrated that a collision between 
the GRB host galaxy and its companion has led to the formation of a dense atomic gas ring in the galaxy \citep[see also][for the presence of a similar ring in the host galaxy 
of famous transient AT2018cow]{Roychowdhury19-2019MNRAS.485L..93R}. Simulations show that such rings have  radial 
expansion velocities  of $\sim$100 \kms\, \citep[][]{Renaud18-2018MNRAS.473..585R}, i.e. traveling  a distance of 300 pc in about 3 Myr. It is thus plausible  that the passage of the ring-shaped density wave triggered gas compression a few  Myrs  
ago at the position of the GRB. This then led to the in situ formation of the clusters and the GRB 
progenitor, and finally the feedback from massive stars in the clusters  
dispersed  the remaining gas. Meanwhile the gas ring moved    outward, leading to the formation of dense clouds outwards 
from the GRB position (C6, C7, C8 and C9). These   clouds are still in the process of assembling, and have not yet 
formed a significant number  of stars, as evident from the long $\tau_{\rm dep}$ in this region. 
{
It is noteworthy that  the \ha\, image   of the galaxy does not show the presence of recent star formation  outside 
the \hi ring while \ha\, emission is  present only inwards of the  gas ring in the galaxy (see the left panel of Figure \ref{fig:muse-alma}), which supports this scenario.
}

The clumps with short depletion times are either in the nucleus of the galaxy (C1) or are in the upstream side of the  gas ring (C4-C5). 
%The short depletion times  found in the C1 and C3-C4    correspond to the downstream side of the ring  and the nucleus (C1). 
The $0.37-1.25$ Gyr upper limit on the depletion time at the location of GRB is also consistent with its correspondence to the 
upstream side of the ring . 
However, the clumps on the downstream side (in this scenario, C6, C7, C8, C9, and C10) yield a significantly longer depletion times  
(note that C10 too is offset by about 0.6\arcs, or 100 pc,  from the stellar component visible in the  HST image). 
In other words, the variation of depletion times in the molecular gas clumps within the atomic gas ring demonstrates the different  
stages of enhancement in  
star formation  triggered by the ring density wave induced by the collision between the GRB host galaxy  and its companion. 
Note that a similar pattern is found in simulations of ring galaxies where the ring 
structure has starburst-like physical conditions \citep[][]{Renaud18-2018MNRAS.473..585R}. 
{Together with the galaxy-scale ring geometry seen in \hi  \citep[see the left panels of Figures \ref{fig:hst-alma} and  \ref{fig:muse-alma}, 
and the detailed discussion in][]{Arabsalmani19-2019MNRAS.485.5411A}, this supports  the idea of a ring-triggered compression of 
the molecular gas which leads to locally enhanced star formation, and of the GRB progenitor having formed in situ. 
}

%C8  appears to be  associated with  the star forming region about  600 pc to   the southwest of GRB 980425, but offset by about 0.5 \arcs ($\sim$ 90 pc) from the stellar component visible in the  HST image. 

\section{Summary}
\label{sec:sum}

We present ALMA observations of molecular gas in  the host galaxy of GRB 980425 (ESO 184--G82) at $z=0.0086$ on  
$\sim$50 pc scales. ESO 184--G82, on galactic scales  has  properties similar  to those of nearby dwarf galaxies and   
does not show any sign of having star formation in the starburst mode. We identify several molecular gas clumps within the galaxy. 
The clump closest to the location of the GRB  is separated by approximately 280 pc from the GRB position. We argue that 
in the unlikely  scenario where the GRB progenitor was a runaway star, it is more likely that it was ejected from a cluster 
formed in this gas clump than from the WR region as previously suggested. However, it seems most likely  that the progenitor was formed 
in situ and that the native cloud of the progenitor was dispersed due to stellar feedback in the region. 

We measure the molecular gas depletion time on cloud scales in ESO 184--G82 and found a  spread of 1.9 dex in  the measured  
$\tau_{\rm dep}$ values. The large spread of 1.9 dex  is comparable with the large variations of $\tau_{\rm dep}$ found in  starburst galaxies.  
This  supports   the idea of a local enhancement of star formation in the galaxy 
triggered by the passage of a ring-shaped density wave due to a  collision between the galaxy and its companion 
discussed in \citet[see][]{Arabsalmani19-2019MNRAS.485.5411A}.  
Our measurements for the $\tau_{\rm dep}$ in ESO 184--G82  therefore indicate the presence of (local) starbursts mode of 
star formation in ESO 184--G82, which is  the favourable condition  for the formation of  dense and massive star clusters.   
We  estimate the   molecular gas masses of the  identified  clumps and found them to be indeed sufficient for  the formation of  massive 
star clusters.

This study  supports  the idea in 
which massive progenitor stars of GRBs are formed in starburst conditions  commonly triggered by interactions at lower redshifts 
($z\lesssim$1.0). In rare cases local starburst conditions are caused due to the internal dynamics of the galaxies, 
as was shown to be the case in the host galaxy of SLSN PTF10tpz \citep[see][]{Arabsalmani19-2019ApJ...882...31A}. 
Extending resolved studies of molecular gas to a larger number of host galaxies of massive star 
explosions at low redshifts is necessary to confirm  the formation of progenitors under such extreme conditions.

%\acknowledgements 
\section*{Acknowledgments}
We would like to thank the referee very much for extremely helpful comments and suggestions which improved the paper enormously. 
M.A. would like to thank Pavel Kroupa and Brent Groves for very helpful discussions. 
M.A. and S.R. acknowledge support from  the Australian Research Council Centre of Excellence for All Sky Astrophysics in 3 Dimensions (ASTRO 3D)  through project number CE170100013. 
F.R. acknowledges support from the Knut and Alice Wallenberg Foundation. 
This project has received funding from the European Research Council (ERC) under the European Union’s Horizon 2020 research and 
innovation programme (grant agreement No. [725246]). 
ALMA is a partnership of ESO (representing its member states), NSF (USA) and 
NINS (Japan), together with NRC (Canada), NSC and ASIAA (Taiwan), and KASI (Republic of Korea), 
in cooperation with the Republic of Chile. The Joint ALMA Observatory is operated by ESO, AUI/NRAO 
and NAOJ. 
We  acknowledge using  data based on observations collected at the European Southern Observatory under ESO programs 064.H-0375(A), 066.D-0576(A),  165.H-0464(A), and  095.D-0172(A). 
This work is based in part on   data made with the NASA/ESA Hubble Space Telescope, obtained from the data archive at the Space Telescope Science Institute.

\end{document}